\documentclass[12pt, reqno]{gsm-l}
\pdfoutput=1
\usepackage{amsmath,amssymb,amsthm}
\usepackage{graphicx}
\usepackage{enumerate}
\usepackage[colorlinks=true]{hyperref}
\usepackage{comment}
\usepackage{youngtab,cancel}
\usepackage[american]{babel}
\usepackage{bm}
\usepackage{feynmf}
\usepackage{latexsym}

\newcommand{\beq}{\begin{equation}}
\newcommand{\eeq}{\end{equation}}
\newcommand{\beqa}{\begin{eqnarray}}
\newcommand{\eeqa}{\end{eqnarray}}
\def\lsim{\ \rlap{\raise 3pt \hbox{$<$}}{\lower 3pt \hbox{$\sim$}}\ }
\def\gsim{\ \rlap{\raise 3pt \hbox{$>$}}{\lower 3pt \hbox{$\sim$}}\ }

\def\fv{\mathbf{5}}
\def\bfv{\mathbf{\bar{5}}}

\def\for{\mathbf{ 45}}
\def\bfor{\mathbf{\bar{45}}}

\def\se{\mathbf{ 70}}
\def\bse{\mathbf{\bar{70}}}

\def\tf{\mathbf{24}}

 \numberwithin{table}{chapter}
 \numberwithin{figure}{chapter}
 \numberwithin{equation}{chapter}
 \numberwithin{section}{chapter}

\newtheorem{defin}{Definition}

\newtheorem{prop}{Proposition}
\newtheorem{veri}{Verification}

\includecomment{notes}
\specialcomment{notes}
{\begingroup}{\endgroup}
\excludecomment{notes}

\def\fv{\mathbf{5}}
\def\bfv{\mathbf{\bar{5}}}

\def\for{\mathbf{ 45}}
\def\bfor{\mathbf{\bar{45}}}
\def\se{\mathbf{ 70}}
\def\bse{\mathbf{\bar{70}}}
\def\tf{\mathbf{24}}

\def\bSigma{\mathbf{\Sigma}}

\begin{document}

\vspace{-5.0cm}		
\title{Phenomenology of supersymmetric models with anomalous symmetries $U(1)_{H}$\\
\textmd{by:}\\
\textmd{Mauricio de Jes\'us Vel\'asquez L\'opez}\\
 \textmd{Under the direction of prof. Diego Restrepo Quintero}\\
\vspace{1.0cm}
\hspace{0.6cm}\includegraphics[scale=0.95]{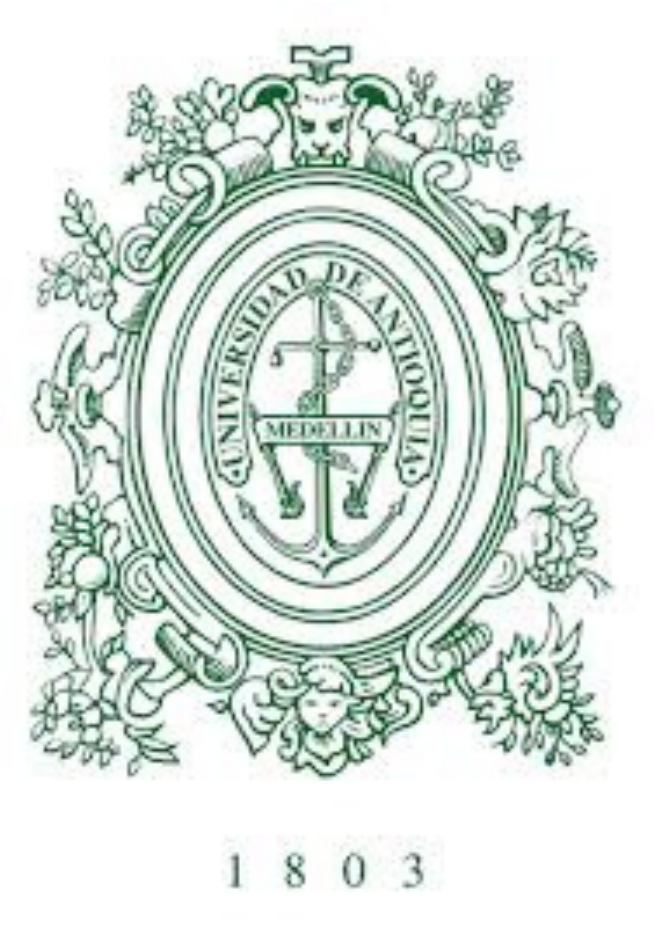}\\
\vspace{1.5cm} 
\textmd{Thesis submitted in partial fulfillment }\\
\textmd{for the degree of Doctor in physics}\\
\textmd{Physics institute}\\
\textmd{University of Antioquia}\\
\textmd{Medell\'in-Colombia}\\
$2013$}
\date{\today}  

\maketitle

\begin{flushright}
		\begin{minipage}[c]{3in}
				{A mis padres, Nelly L\'opez Rold\'an y Emilio Vel\'asquez Garc\'ia,  con gratitud y cari\~no por la interacci\'on que me permiti\'o conocer el mundo. }
		\end{minipage}
\end{flushright}
\vspace{11.0cm}

\newpage\thispagestyle{empty}

 \setcounter{page}{1}
\tableofcontents
\listoftables
\listoffigures
\chapter{Introduction}

\begin{flushright}
		\begin{minipage}[c]{3in}
				\tiny\bfseries\emph{"We must not seek, but find, we must not judge, but observe and comprehend, inspire and elaborate the inspired. We have to feel our own essence integrated and ordered at whole. Only then we will have real relations with nature."\\ \\
                     \quad Hermann Hesse}
		\end{minipage}
\end{flushright}
\vspace{1.5cm}

The standard model (SM) is a very successful framework for describing
particle physics phenomena. However, it suffers from some serious
phenomenological problems, among which: neutrinos are massless, the
conditions for baryogenesis are not fulfilled, and there is no
candidate for dark matter (DM).  The first two problems can be
solved by extending the SM to include the seesaw mechanism for
neutrino masses~\cite{Minkowski:1977sc,Yanagida:1979as,Glashow,GellMann:1980vs,Mohapatra:1980yp} 
that also opens the possibility of
baryogenesis via leptogenesis~\cite{Fukugita:1986hr,Davidson:2008bu},
while extending the SM to its supersymmetric version (SSM) can provide
a natural candidate for DM.  However, in contrast to the SM, the SSM
does not have accidental lepton ($L$) and baryon-number ($B$)
symmetries, and this can lead to major phenomenological problems, like
fast proton decay.  The standard solution to forbid all dangerous
operators is to impose a discrete symmetry, $R$--parity, and only in
the $R$-parity conserving SSM the lightest supersymmetric particle (LSP),
generally  the neutralino, is stable, and provides a good DM
candidate.

Similarly to the SM, the SSM does not provide any explanation for
the strong hierarchy in the charged fermion Yukawa couplings. One way to
explain the flavor puzzle and the suppression of the fermion masses
with respect to the electroweak breaking scale is to impose Abelian
flavor symmetries, that we generically denote as $U(1)_H$, that are
broken by SM-singlets commonly denoted as flavons. This process involves horizontal charges for the fields that determines whether a particular term can or cannot be present in the superpotential.
After this problem is solved remain some free horizontal charges,
that can be used to set the order of magnitude of the $R$-parity violating
couplings. In supersymmetric models extended to include an anomalous horizontal
$U(1)_H$ symmetry {\it \`a la} Froggatt-Nielsen (FN) \cite{Froggatt:1978nt},
the standard model particles and their superpartners do not carry a
$R$-parity quantum number and instead carry a horizontal charge
($H$--charge). For a review see \cite{Dreiner:2003hw}.  In addition, these
kinds of models involve new heavy FN fields and, in the simplest
realizations, an electroweak singlet superfield $S$ of $H$--charge $-1$.
$R$-parity conserving as well as $R$-parity violating $SU(3)_{C}\times
SU(2)_{L}\times U(1)_Y\times U(1)_H$--invariant effective terms arise once
below the FN fields scale, $M$, the heavy degrees of freedom are
integrated out. These terms involve factors of the type $(S/M)^n$,
where $n$ is fixed by the horizontal charges of the fields implicated
and determines whether a particular term can or cannot be present in the
superpotential. The holomorphy of the superpotential forbids all the
terms for which $n<0$ and although they will be generated after
$U(1)_H$ symmetry breaking (triggered by the vacuum expectation value of the scalar
component of $S$, $\langle S\rangle$) via the K\"ahler potential
\cite{Giudice:1988yz} these terms are in general much more suppressed than
those for which $n\ge0$.  Terms with fractional $n$ are also forbidden
and in contrast to those with $n<0$ there is no mechanism through
which they can be generated.  Finally, once $U(1)_H$ is broken the
terms with positive $n$ yield Yukawa couplings determined---up to order
one factors---by $\theta^n=(\langle S\rangle/M)^n$. The standard
model fermion Yukawa couplings typically arise from terms of this
kind. Correspondingly, supersymmetric models based on an $U(1)_H$
Abelian factor are completely specified in terms of the $H$--charges. 
Then the $R$-parity conservation can be for example enforced by
an extended gauge symmetry together with supersymmetry (that requires
a holomorphic superpotential) as in the model studied
in~\cite{Mira:1999fx}, or solely by the gauge symmetry thanks to a
suitable choice of the $U(1)_H$--charges, as in
Ref.~\cite{Dreiner:2003yr}.

However, for scenarios such as the $R$--parity conserving constrained minimal supersymmetric standard model (CMSSM), the recent results on searches for supersymmetry by 
CMS~\cite{Chatrchyan:2012lia} and ATLAS~\cite{Aad:2012fqa}
experiments have raised the bound on scalar and gluino masses, when
they are approximately equal, to the order of 1.4 TeV.
These searches are mainly based on missing transverse momentum carried
by the LSP.
A high mass scale for scalars and gluinos represents a potential chink
in the initial proposal of the SSM as a possible solution to the
hierarchy problem. These mass limits can be avoided in alternative supersymmetric~
models such as the $R$-parity violating SSM
\cite{Hall:1983id,Ross:1984yg,Barger:1989rk,Dreiner:1997uz,Allanach:2003eb,Barbier:2004ez},
in which the LSP is usually assumed to be the gravitino that also
provides a good decaying dark matter candidate
~\cite{Takayama:2000uz,Buchmuller:2007ui}.
The next-to-the-lightest supersymmetric particle decays to standard
model particles, and thus the missing transverse momentum may be
considerably reduced
~\cite{deCampos:2007bn,Brust:2011tb,Butterworth:2009qa,Allanach:2012vj,Brust:2012uf,Asano:2012gj,Curtin:2012rm,Graham:2012th}. 
In addition, if the involved couplings are small enough, the presence of
displaced vertices may reduce the efficiency of the standard searches
at the LHC~\cite{deCampos:2007bn,Graham:2012th}. The simplest case of an anomalous horizontal symmetry with a single flavon, can also suppress, but do not completely prohibit, $R$--parity violating terms. Along these lines, consistent models have been built in which neutrino oscillation data can be explained\cite{ Mira:2000gg,Dreiner:2006xw, Dreiner:2003hw,Dreiner:2003yr,Choi:1998wc,Dreiner:2007vp}. 
Also, by using the reported anomalies in cosmic-ray electron/positron
fluxes, a consistent model with tiny $R$-parity breaking couplings was
built with decaying leptophilic-neutralino dark
matter~\cite{Sierra:2009zq}.

We adopt in this thesis a new approach by assuming a set of $H$-charges that
give rise to a self-consistent model of $R$-parity breaking and
baryon-number violation.
As a consequence of our $H$-charge assignments, it is not
possible to generate a Majorana mass term for left-handed neutrinos. 
However, a neutrino Dirac matrix can be built after the introduction of
right-handed neutrinos with proper $H$-charges.
We also show that by adding a second flavon field with fractional
charge, it is possible to build a Majorana neutrino mass matrix.
In both cases an anarchical matrix
~\cite{Hall:1999sn,Haba:2000be,deGouvea:2003xe,deGouvea:2012ac} is
obtained, which is supported by the recent results of a large value
for $\theta_{13}$~\cite{Abe:2011sj,Abe:2011fz,An:2012eh,Ahn:2012nd}.

As a consequence of $H$-charge assignments, the $\lambda''_{323}$
coupling dominates over the other couplings, and the third-generation
quarks are expected to be present at the final states of LSP decays.  Moreover,
the horizontal symmetry predicts a precise hierarchy of $B$-violating
couplings, which can be translated into relations between different
branching ratios, that could be measured at $e^+\ e^-$ colliders. The required conditions to obtain one
$R$-parity breaking SSM with $B$ violation are shown, also taking into account
dimension-five operators.

Next, and continuing in the structure of the FN mechanism extended with a horizontal symmetry $U(1)_H$, we introduce the model $SU(5)$ proposed by
Georgi y Glashow~\cite{Georgi:1974sy}, which incorporates the standard model group and gives a description in terms of a single constant $g_{5}$; moreover, the quantization of charge comes as a direct consequence of the algebra of $SU(5)$, and the lifetime of the proton is consistent with the current experimental bounds~\cite{Abe:2013lua}. Differently from the SM case~\cite{Dreiner:2003yr}, in $SU(5)$ GUTs it
is rather difficult to implement this kind of horizontal symmetries,
because there is less freedom in choosing the $H$--charges (see for
example~\cite{Chen:2008tc}). However, if the flavons that break the
horizontal symmetry are assigned to the adjoint representation of
$SU(5)$~\cite{Aristizabal:2003zn,Duque:2008ah,Wang:2011ub}, charges
that were forbidden in the singlet flavon case become allowed, under
the assumption that certain representations for the Froggatt-Nielsen
(FN)~\cite{Froggatt:1978nt} messengers fields do not exist.  In
contrast to the non-unified $SU(3)_{C}\times SU(2)_{L}\times U(1)_{Y}\times
U(1)_H$ model, where the singlet nature of the flavons is mandatory,
in $SU(5)\times U(1)_H$ assigning the flavons to the adjoint has the
additional bonus that non-trivial group theoretical coefficients
concur to determine the coefficients of the effective
operators~\cite{Aristizabal:2003zn,Duque:2008ah,Wang:2011ub}.  In this
case, under the additional assumption that at the fundamental level
all the Yukawa couplings obey to some principle of
universality~\cite{Duque:2008ah}. A virtue of the $U(1)_H$ gauge symmetry
implemented here is that when the $U(1)_H$ charges are chosen
appropriately, $\Delta B\neq 0$ and $\Delta L=1$ operators are
forbidden at all orders.  However, $\Delta L=2$ operators
corresponding to Majorana masses for heavy neutral fermions of the
seesaw remain allowed, and thus the seesaw mechanism can be embedded
in the model.  More in detail, following~\cite{Dreiner:2003yr} we
chose the $H$-charges in such a way that operators with even
$R$--parity have an overall $H$-charge that is an integer multiple of
the charge of the $U(1)_H$ breaking scalar fields (that, without loss
of generality, we set equal to $\pm 1$).  In contrast, all the
$R$--parity breaking operators, that have an overall half-odd-integer
$H$--charge, are forbidden. Then, to allow for $\Delta L=2$ Majorana
masses while forbidding $\Delta L=1$ operators, it is sufficient to
chose the $H$--charges of the heavy seesaw neutral states as
half--odd-integers. The order one coefficients that
determine quantitatively the structure of the mass matrices become
calculable.

The structure of the thesis is as follows:

Chapter~\ref{cap1}, Sec.~\ref{sec:BNVmodel22}, will be an introduction to the theory of FN~\cite{Froggatt:1978nt} to explain the mass spectrum in the fermionic sector.
In Sec.~\ref{modteo} we enunciate the selection rules for our model.
In Sec.~\ref{gag} we show how $R$--parity can be obtained from a gauge symmetry. Choosing a suitable set of horizontal charges, the
$R$ parity comes as a direct result of this election.
~\cite{Dreiner:2003yr}. In Sec.~\ref{cams} we find the $U(1)_{H}$--charges for the fields of the SSM in terms of 4 four parameters. In Sec.~\ref{theta} we calculate the numerical value if the expansion parameter is $\theta$. In Sec.~\ref{modelt} we made ​​the calculations of $U (1)_H$--charges for the $R$--parity breaking terms. In Sec.~\ref{unif} we synthesize the problem with several flavons for the unification theory $SU(5)$.

Chapter \ref{cap2}, Sec.~\ref{sec:BNVmodel}, will be raised the conditions to obtain one
$R$-parity breaking SSM with $B$ violation, also taking into account
dimension-five operators.
The generation of neutrino masses by introducing right-handed
neutrinos is discussed in 
Sec.~\ref{sec:RHN}. In Sec.~\ref{sec:implications}
the consequences for collider physics are mentioned.

Chapter~\ref{cap3} will present a $SU(5)\times U(1)_H$ supersymmetric model for neutrino
masses and mixings.

In Chapter~\ref{sec:conlusions} are presented the discussions and conclusions of the work.

In the Appendix ~\ref{aped.B} are presented the horizontal charges of dimension-five $R$-parity breaking operators in detail while in~\ref{aped.C} we calculate explicitly a contribution to  ${Y}_\nu=\sum_i {Y}^{(i)}_\nu$ 
at $\mathcal{ O}(\epsilon)$ of $SU(5)$.

\chapter{Supersymmetric models with a symmetry $U(1)_H$}\label{cap1}

One of the unsolved problems in the standard model is the
related to the hierarchy of the masses, in charged fermion sector.
The central idea proposed by FN 
~\cite{Froggatt:1978nt,Dreiner:2003hw} was to introduce an Abelian $U (1)_{H}$--symmetry which assigns $H$ charges to fermion fields to try to explain this mass hierarchy. 
These masses may be expressed in terms of a parameter $\theta$
given by:
\begin{equation}\label{1}
    \theta=\frac{\langle S \rangle}{M_{F}},
\end{equation}
where $S$ is a flavonic field, and $M_{F}$ is a mass scale for the FN heavy Fields.
\section{Froggatt-Nielsen mechanism in the quark sector}
\label{sec:BNVmodel22}
\begin{figure}
\centering
\includegraphics[scale=0.8]{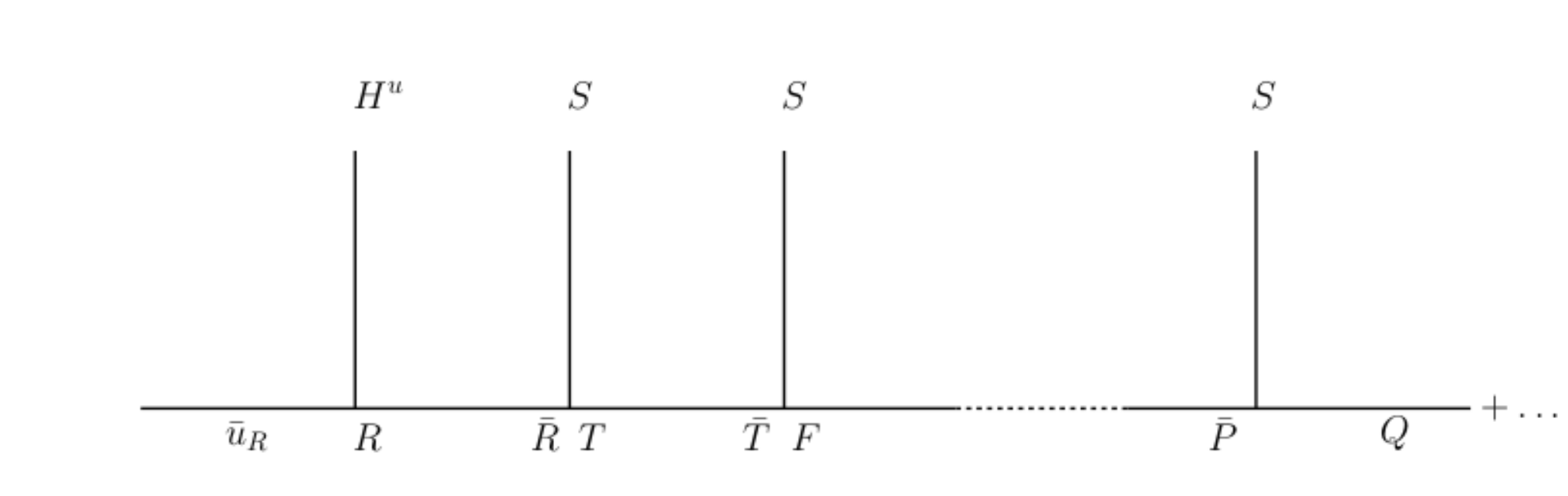}
\caption{Diagrammatic representation of the Yukawa Lagrangian, where are taken $n$ Froggatt-Nielsen fields that allow the invariance under $U(1)_{H}$. } 
\label{fig:FN}
\end{figure}

The Yukawa Lagrangian in the standard model includes terms of
type:
\begin{equation}\label{2}
    \mathcal{L}=h^{u}\bar{u}_{R}H^{u}Q + \text{h.c},
\end{equation}
where $Q=(u,d)_{L}$ transforms as a representation $(\mathbf{3},\mathbf{2})_{1/3}$ under $SU(3)_{C}\times SU(2)_{L} \times U(1)_{Y}$;
$\bar{u}_{R},\bar{d}_{R}$ transforms as a $(\mathbf{\bar{3}},\mathbf{1})_{-4/3, 2/3}$ and
$H^{u}$ transforms as $(\mathbf{1}, \mathbf{\bar{2}})_1$. The Lagrangian above the scale
$M_{F}$ is:
\begin{equation}\label{3}
  \mathcal{L}= \bar{u}_{R}H^{u}R + \bar{R}ST + \bar{T}SF +
  \ldots + \bar{P}SQ + \text{h.c},
\end{equation}
where $S$ is a flavon which transforms as a singlet under $SU(3)_{C}\times SU(2)_{L} \times U(1)_{Y}$ and acquires a vacuum expectation value given by $\langle S \rangle$ a scale below
$M_{F}$. $R$, $T$, $F$, $P$ are heavy fields, charged under the symmetry $U(1)_{H}$. The diagrammatic representation of Eq.~(\ref{3}) 
 is shown in Figure~\ref{fig:FN}. This Feynman diagram will lead to effective contributions of the mass terms of the fermions when the Abelian $U(1)_{H}$ symmetry gets spontaneously broken. This diagram must be  invariant under the horizontal $U(1)-H$--charge assignment.
The horizontal charges are defined as:
\begin{equation}\label{4}
S=H(S),\quad Q=H(Q), \quad H^{u}=H(H^{u}), \quad
u=H(\bar{u}_{R}),\quad d=H(\bar{d}_{R}), \quad
\end{equation}
and
\begin{equation}\label{5}
R=H(R), \quad T=H(T), \quad F=H(F),\quad
\ldots, P=H(P).
\end{equation}
As the horizontal $U(1)_{H}$--charge must be conserved in each of the vertices of Figure~\ref{fig:FN}, we have
\begin{equation}\label{6}\nonumber
   u + H^{u} + R=0,
\end{equation}
\begin{equation}\label{7}\nonumber
  -R + S + T=0,
\end{equation}
\begin{equation}\label{8}\nonumber
  -T + S + F=0,
\end{equation}
\quad\quad\quad\quad\quad \quad \quad  \quad\quad \quad \quad \quad \quad \quad\quad\quad\quad
\;\vdots\quad \quad \quad\quad\vdots \quad\quad \quad\vdots
\begin{equation}\nonumber
  -P + S + Q=0,
\end{equation}
by integrating the heavy fields of Froggatt-Nielsen in the Eq.~(\ref{3}), are added the $n$ charges $U(1)_{H}$ assigned to these fields, thereby obtaining:
\begin{equation}\label{11}
 \cancel{R}-\cancel{R} + S + \cancel{T} - \cancel{T} +
S + \cancel{F} +  \ldots - \cancel{P} + S=
n S=n.
\end{equation}
Therefore, after the breaking of the horizontal $U(1)_{H}$--symmetry and the integration of heavy fields, the effective Lagrangian in the Eq.~(\ref{3}) is:
\begin{equation}\label{10}
    \mathcal{L}^{\text{Yukawa}}=\bar{u}H^{u}\left(\frac{\langle S
    \rangle}{M_{F}}\right)^{n}Q + \text{h.c}.
\end{equation}
For this term be invariant under the symmetry $U(1)_{H}$, it must to comply with:
\begin{equation}\label{12}
n=u + H^{u} + Q,
\end{equation}
where the Yukawas are:
\begin{align}\label{15}
    h^{u}&\sim \theta^{n}\sim \theta^{u + H^{u} +Q}
\end{align}
The assignment of an appropriate set of horizontal $U(1)_{H}$--charges
 for these fields could give a model
phenomenologically viable with experimental observations.

To summarize, one can say that the idea proposed by FN consists of introducing an Abelian horizontal symmetry
$U(1)_{H}$, and some scalar field $S$, called flavon, together with the need to assume that there are a large number of heavy FN fields that serve as mediators of new interactions.
 These heavy fields are vectorlike. With this set of conditions the suppression of Yukawa couplings can be explained.

\section{General renormalizable superpotential}\label{modteo}
 The most general renormalizable superpotential including right--handed neutrinos, is given by~\cite{Dreiner:2003hw}:
\begin{align}\label{po}
   \mathcal{W}&=
   \varepsilon^{ab}\delta^{xy}h^{u}_{ij}\widehat{Q}^{i}_{xa}\widehat{H}^{u}_{b}\widehat{u}^{j}_{y}\nonumber
   +
   \varepsilon^{ab}\delta^{xy}h^{d}_{ij}\widehat{Q}^{i}_{xa}\widehat{H}^{d}_{b}\widehat{d}^{j}_{y}\\ \nonumber
   &+   \varepsilon^{ab}h^{l}_{ij}\widehat{L}^{i}_{a}\widehat{H}^{d}_{b}  \widehat{l}^{j} +
   \varepsilon^{ab}Y^{\nu}_{ij}\widehat{L}^{i}_{a}\widehat{H}^{u}_{b}\widehat{N}^{j}\\ \nonumber
   &+ \varepsilon^{ab}\mu \widehat{H}^{d}_{a}\widehat{H}^{u}_{b} +
   M^{\text{R}}_{ij}\widehat{N}^{i}\widehat{N}^{j}\\\\ \nonumber \nonumber
   &+ \frac{1}{2}\varepsilon^{ab}\lambda_{ijk}\widehat{L}^{i}_{a}\widehat{L}^{j}_{b} \widehat{l}^{k}\\ \nonumber
   &+
   \varepsilon^{ab}\delta^{xy}\lambda^{'}_{ijk}\widehat{Q}^{i}_{xa}\widehat{L}^{j}_{b}\widehat{d}^{k}_{y}\nonumber
   + \Xi_{i}\widehat{N}^{i}\\ \nonumber
   &+
   \frac{1}{2}\varepsilon^{xyz}\lambda_{ijk}^{''}\widehat{u}^{i}_{x}\widehat{d}^{j}_{y}\widehat{d}^{k}_{z}\nonumber
   + \varepsilon^{ab}Y^{\nu}_{i}\widehat{N}^{i}\widehat{H}^{d}_{a}\widehat{H}^{u}_{b}\\ \nonumber
   &+ \varepsilon^{ab}\mu^{i}\widehat{L}^{i}_{a}\widehat{H}^{u}_{b} + y^{\text{R}}_{ijk}\widehat{N}^{i}\widehat{N}^{j}\widehat{N}^{k},\nonumber
\end{align}
the upper block in Eq.~(\ref{po}) is $R$--parity conserving, the lower block violates $R$--parity. In the Eq.~(\ref{po}) $\widehat{H}$, $\widehat{Q}$, $\widehat{L}$ represent the left--chiral $SU(2)_{L}$--doublet superfields of the higgses, the quarks and leptons; $\widehat{u}$, $\widehat{d}$, $\widehat{l}$, $\widehat{N}$ represent the right--chiral superfields; $a$, $b$, $c$ and $x$, $y$, $z$ are $SU(2)_{L}$-- and $SU(3)_{C}$--indices, $i$, $j$, $k$ are generational indices; $\delta^{xy}$ is the Kronecker symbol, $\varepsilon^{\ldots}$ symbolizes any tensor that is totally antisymmetric with respect to the exchange of any two indices, with $\varepsilon^{12\ldots}=1$. All other symbols are coupling constants, The $\lambda_{ijk}$ ($\lambda^{''}_{ijk}$) being antisymmetric with respect to the exchange of the first two (last two) indices. Here, the simultaneous presence of terms that violate Baryon number ($B$) and lepton number ($L$) give a very short proton lifetime (For a more detailed explanation, see~\cite{Martin:1997ns}). 
Lepton number is explicitly broken by the bilinear couplings $\mu_i$
and trilinear couplings $\lambda_{ijk}$ and $\lambda_{ijk}'$, whereas
the couplings $\lambda_{ijk}''$ are responsible for the $B$ violation.
The factor of $1/2$ is due to the antisymmetry of the corresponding
operators~\cite{Barbier:2004ez}.
The $H$ charge for the fields determines whether or not a particular term can be present in the
superpotential. As will be seen in the next section, when extending a supersymmetric model with a
$U(1)_H$ Abelian factor, the size of all the parameters entering in the
superpotential arises as a consequence of $U(1)_H$ breaking. In
particular, the violating lepton or baryon--number may be absent without the need of
$R$--parity~\cite{Mira:2000gg,Dreiner:2006xw, Dreiner:2003hw,Dreiner:2003yr, Choi:1998wc, Dreiner:2007vp,Joshipura:2000sn}.
Before proceeding we will fix our notation: following Ref.~\cite{Mira:2000gg}, we will denote a field and its $H$ charge with the same symbol, i.e. $H(f_{i})=f_{i}$, $H$--charge differences as $H(f_i-f_j)=f_{ij}$~\cite{Dudas:1995yu}; bilinear $H$ charges as $n_{i}=L_{i}+H_{u}$. In
what follows we will constrain the $H$--charges to satisfy the condition
$|H(f_i)|\lesssim10$ that as highlighted in Refs.~\cite{Dreiner:2003hw,Choi:1996se} leads to a complete
consistent supersymmetric flavor model. Trilinear
$H$--charges of the $B$ and $L$ violating operators will be written as $n_{\lambda_{ijk}}$ with the index determined
by the corresponding trilinear coupling, that is to say the index can
be given by $\lambda_{ijk}$, $\lambda_{ijk}'$, or $\lambda_{ijk}''$.
We fix $\theta=\langle S \rangle/M_{\text{P}}\simeq 0.22$~\cite{Dreiner:2003yr,Irges:1998ax} and $S=-1$.
The holomorphy of the superpotential forbids all the
terms for which $n<0$ (where $n$ is an abbreviation for the overall $H$--charge of an operator in the superpotential) and although they will be generated after
$U(1)_H$ symmetry breaking (triggered by the vacuum expectation value of the scalar
component of $S$, $\langle S \rangle$) via the K\"ahler potential
~\cite{Giudice:1988yz} these terms are in general much more suppressed than
those for which $n\ge0$.  Terms with fractional $n$ are also forbidden
and in contrast to those with $n<0$ there is no mechanism through
which they can be generated.
As already stressed any coupling in the superpotential is determined
up to order 1 factors by its $H$--charge. Thus, any bilinear or
trilinear couplings $\mu_i$ and $\lambda_T$ (where $\lambda_T$ is an abbreviation for any of the trilinear couplings in Eq.~(\ref{po})) must be given by
~\cite{Dreiner:2003hw,Binetruy:1996xk}
\begin{align}
  \label{eq:res}
  \mu_\alpha\sim &
  \begin{cases}
    M_P\theta^{n_\alpha}         & n_\alpha\ge 0\\
    m_{3/2}\theta^{|n_\alpha|} & n_\alpha<0\\
    0                        & n_\alpha\ \text{fractional}
  \end{cases}
  &\lambda_T\sim&
  \begin{cases}
    \theta^{n_\lambda}               & n_\lambda\ge 0\\
    (m_{3/2}/M_P)\theta^{|n_\lambda|} & n_\lambda<0\\
    0                               &n_\lambda\ \text{fractional}
  \end{cases}\,.
\end{align}

An operator with fractional charges is prohibited also on models with several flavons of integer charges.
\section{$R$--parity as a result of a gauge symmetry}\label{gag}

It can be shown that the conservation of $R$--parity in the MSSM,
may result as a consequence of the proper choice of the horizontal $U(1)_{H}$--charges. 

Now, the overall $U(1)_{H}$--charge for any operator can be written as
follows:
\begin{align}\label{25}
H_{\text{Total}}&=\sum_{i}(n_{N^{i}}N^{i}) +
\sum_{i}(n_{L^{i}}L^{i} + n_{l^{i}}l^{i}) +
n_{H^{d}}H^{d} + n_{H^{u}}H^{u} \\ \nonumber
&+
\sum_{i}(n_{Q^{i}}Q^{i}
 + n_{d^{i}}d^{i}+
n_{u^{i}}u^{i}), \nonumber
\end{align}
where the $n_{N^{i}}, n_{L^{i}},  \ldots$ are positive integers representing the number of times it is repeated a field. These numbers are independent, due to the gauge invariance of the group
$SU(3)_{C}\times SU(2)_{L}\times U(1)_{Y}$, moreover we define in Eq.~(\ref{25}):
$n_{Q}=\sum_{i}n_{Q^{i}}$, $n_{L}=\sum_{i}n_{L^{i}}$, $n_{u}=\sum_{i}n_{u^{i}}$, $\ldots$
The following example shows how to find the overall
$U(1)_{H}$--charge from some operator of the Eq.~(\ref{po}). For the term
$h^{u}QH^{u}u$
, we obtain:
 \begin{align*}
    n_{Q}&=1,\\
   n_{H^{u}}&=1,\\
      n_{u}&=1,\\
\end{align*}
The total charge for this operator is:
\begin{align*}
    H_{\text{Total}}= Q+ H^{u} + u.
\end{align*}
In the same way are calculated the overall $U (1)_{H}$--charges for
other operators in the Eq.~(\ref{po}). Below, 
are listed two central definitions, which form the main structure for the following analysis:
\begin{defin}\label{def1}
For the different fields in the model, we may use:
\begin{align}\label{26}
    B_{p}&=(-1)^{n_{Q} - n_{u} - n_{d}},\\ \nonumber
    L_{p}&=(-1)^{n_{L} - n_{N} - n_{l}},\\ \nonumber
    R_{p}&= B_{p}\times  L_{p}= (-1)^{n_{Q} - n_{u} - n_{d}+ n_{L} - n_{N} - n_{l}} \nonumber
\end{align}
Where the $B_{p}$, $L_{p}$, $R_{p}$ correspond to baryonic parity, leptonic parity and $R$--parity. All operators,
which conserve the $\mathbb{Z}_{2}$--symmetry  $R_{p}$, each have an overall integer
$H_{Total}$--charge. Namely, all operators for which
     $n_{Q} - n_{u} - n_{d}+n_{L} - n_{N} -
    n_{l}$ is even, each have an overall integer $U(1)_{H}$--charge.
\end{defin}

\begin{defin}\label{def2}
One operator which violates $R_p$ have an overall fractional $U(1)_{H}$--charge. Namely, one operator for which $n_{Q} -
n_{u} - n_{d}+n_{L} - n_{N} -
    n_{l}$ is odd each have an overall fractional $U(1)_{H}$--charge.
\end{defin}

We can draw several conclusions:

\begin{prop}
 If the field $Q^{1}$ has the same quantum numbers as
    $Q^{2}$, an $R_{p}$--conserving operator $Q^{1}\phi_{1}\phi_{2}\ldots \phi_{n}$ guarantees that $Q^{2}\phi_{1}\phi_{2}\ldots
    \phi_{n}$ is $R_{p}$--conserving as well, we thus find that it is necessary that the charge 
    $Q^{2} - Q^{1}$ is integer.
\end{prop}
\begin{veri}
We start from the Eq.~(\ref{25}). The overall $U(1)_{H}$--charge is calculated 
for this pair of operators,
\begin{align*}
    H^{1}_{\text{Total}}&= n_{Q^{1}}Q^{1} + n_{\phi_{1}}\phi_{1} +
    n_{\phi_{2}}\phi_{2} + \ldots +
    n_{\phi_{n}}\phi_{n}=Z_{1},\\
H^{2}_{\text{Total}}&= n_{Q^{2}}Q^{2} + n_{\phi_{1}}\phi_{1} +
    n_{\phi_{2}}\phi_{2} + \ldots +
    n_{\phi_{n}}\phi_{n}=Z_{2}.
\end{align*}
We subtract this pair of expressions:
\begin{align}\label{prim}
    H^{2}_{\text{Total}} - H^{1}_{\text{Total}}=Q^{2} - Q^{1}= Z_{2} - Z_{1}=
   \text{Integer}.
\end{align}
\end{veri}

\begin{prop}
 For any $SU(3)_{C}\times
 SU(2)_{L}\times U(1)_{Y}$ invariant operator $\phi_{1}\phi_{2}\ldots
    \phi_{n}$ which violates $R_{p}$ one has that $\phi_{1}\phi_{2}\ldots
    \phi_{n}\phi_{1}\phi_{2}\ldots
    \phi_{n}$  conserves $R_{p}$. It follows that all operators which violate $R$--parity 
    have an overall half-odd integer $U(1)_{H}$--charge.
\end{prop}
\begin{veri}
Be $\phi_{1}\phi_{2}\ldots
    \phi_{n}$ an operator which violates $R_{p}$. Using the
    def.~\ref{def2} we find the $H_{\text{Total}}$--charge:
\begin{align}\label{27}
    \phi_{1} +  \phi_{2}+ \ldots +  \phi_{n}=\frac{p}{q}.\quad \textrm{Where $p$ and $q$ belong to integers
    $\mathbb{Z}$}.
\end{align}
Now be the operator $\phi_{1}\phi_{2}\ldots
    \phi_{n}\phi_{1}\phi_{2}\ldots
    \phi_{n}$ that conserves $R$--parity.
    Considering the def.~\ref{def1}, the $H_{\text{Total}}$--charge  of this operator is an integer.
Considering this result, this operator has a integer
$H_{\text{Total}}$--charge of the form:
\begin{align}\label{28}
    2\phi_{1} +  2\phi_{2}+ \ldots +  2\phi_{n}=z ,\; \textrm{where $z$  belong to integers $\mathbb{Z}$}
\end{align}
Introducing the Eq.~(\ref{27}) in the Eq.~(\ref{28}), it is obtained that:
\begin{align*}
    2\frac{p}{q}=z.
\end{align*}
The number $p$ is an odd integer, as $z$ is an integer, it follows that
 $q=2$.
From this it is follows that the $H_{\text{Total}}$--charge for an operator that violates $R$--parity is an half-odd integer number
 $p/2$.
\end{veri}
It follows immediately from the previous preposition and the last terms of the first block of the Eq.~(\ref{po}) that $\widehat{N}$ is half-odd-integer.

\begin{prop}
Let $H^{d}\phi_{1}\phi_{2}\ldots \phi_{n}$ be $SU(3)_C \times SU(2)_L \times U(1)_Y$ invariant and it conserve $R_p$, it follows that $L^{i}\phi_{1}\phi_{2}\ldots \phi_{n}$ does not conserve $R_p$ then the charge $L^{i} - H^{d}$ is half-odd-integer.
\end{prop}
\begin{veri}
Be $H^{d}\phi_{1}\phi_{2}\ldots
    \phi_{n}$ an operator which conserves $R_{p}$.
Given this result, this operator has a 
integer $H_{\text{Total}}$--charge of the form:
\begin{align}\label{29}
    H^{d}+\phi_{1} +  \phi_{2}+ \ldots +  \phi_{n}=p.\quad \textrm{Where $p$ belong to integers
    $\mathbb{Z}$}.
\end{align}
Now be the operator $L^{i}\phi_{1}\phi_{2}\ldots
    \phi_{n}$ that violates $R$--parity.
Considering this result, this operator has a fractional
$H_{\text{Total}}$--charge of the form:
\begin{align}\label{30}
    L^{i}+\phi_{1} +  \phi_{2}+ \ldots +  \phi_{n}=k/2 \; \textrm{Rational number}
\end{align}
We subtract the equation Eq.~(\ref{29}) of the Eq.~(\ref{30}), we get:
\begin{align}\label{31}
    L^{i}-H^{d}=k/2-p=\textrm{half-odd-integer}.
\end{align}
\end{veri}

The MSSM $+\widehat{N}$ $R_p$--conserving, see
~\cite{Dreiner:2003hw}, is of the form:
\begin{align}\label{32}
     Q^{1} +
   H^{d} +
   d^{1}&=n_{1},\\ \nonumber
    Q^{1} +
   H^{u} +
   u^{1}&=n_{2},\\ \nonumber
    L^{1} +
   H^{d} +
   l^{1}&=n_{3},\\ \nonumber
    L^{1} +
   H^{u} +
   N^{i}&=n_{4},\\ \nonumber
   N^{i} +
   N^{j}&=n_{5},
\end{align}
where the $n_{1}, \ldots, n_{5}$ are integers; and analogously for the other matter superfields.

From Eq.~(\ref{26}) one sees that

\begin{align}\label{33}
n_L-n_{N}-n_{l}+n_Q-n_{u}-n_{d}=2\mathcal{R}-\rho,
\end{align}
$\mathcal{R}$ is an integer, $\rho$ is 0 or 1 if $R_p$ is conserved or broken.

We now plug Eqs.~(\ref{prim}),~(\ref{31}),~(\ref{32}),~(\ref{33}) and the Eq.~(3.3) of Ref.~\cite{Dreiner:2003yr} into Eq.~(\ref{25}) and we obtain:
\begin{align}\label{34}
3Q^{1}+L^{1}= \;\textrm{Integer},
\end{align}
is the necessary and sufficient condition ( apart from Eqs.~(\ref{prim}),~(\ref{31}),~(\ref{32})) on the $H$-charges for conserved $R_p$.

\section{The Standard Model fields $H$--charges}\label{cams}

The individual $H$ charges for the SM fields are determined through a set of phenomenological and theoretical conditions.

Eight phenomenological constraints arising from six mass ratios for the quarks and the charged leptons plus the two quark mixing angles
\begin{align} 
    \label{condi1}
    m_u:m_c:m_t &\simeq \theta^{\,8}:\theta^{\,4}:1\,, \nonumber \\
    m_d:m_s:m_b &\simeq
    \theta^{\,4}:\theta^{\,2}:1\,,  \nonumber \\
    m_e:m_\mu:m_\tau &\simeq
    \theta^{\,5}:\theta^{\,2} :1\,,   \nonumber \\
    V_{us}\simeq \theta\,, & \quad  V_{cb}\simeq \theta^{\,2},\,
  \end{align}
where $\theta$, given by the Eq.~(\ref{1}), is a small parameter of the order of the Cabbibo angle $\theta\simeq 0.22$. These eight conditions on the fermion charges can be re-expressed in terms of the following sets of eight charge differences shown in Table~\ref{tab:chargediff}~\cite{Dudas:1995yu, Choi:1996se, Binetruy:1996xk, Binetruy:1994ru, Leurer:1992wg}.
We will not repeat here the phenomenological analysis leading to these sets of charge differences, since this has been extensively
discussed in the literature~\cite{Dudas:1995yu, Choi:1996se, Binetruy:1996xk, Binetruy:1994ru, Leurer:1992wg}. The negative charge differences shown in the reference~\cite{Mira:2000gg} reproduce the matrix elements $V_{ub}$ and $V_{cb}$ much smaller to the observed ones and cannot be improved by the K\"ahler contributions~\cite{Espinosa:2004ya, King:2004tx}. Therefore these charges are not phenomenologically viable.

Two relations are provided by the absolute value of the masses of third generation fermions
\begin{align}\label{condi2}
m_t\simeq\langle H_u \rangle \quad\text{and}\quad m_b\simeq m_\tau.
\end{align}

Two theoretical constraint corresponding to the consistency conditions for the coefficients of the mixed linear anomalies (the second constraint fixes $k_1=5/3$)~\cite{Mira:2000gg, Binetruy:1994ru,Dine:2012mf}
\begin{align}\label{condi4}
\mathcal{A}_{CCH}=\mathcal{A}_{LLH}=\frac{\mathcal{A}_{YYH}}{k_1}=\delta_{\text{GS}},
\end{align}
where the $\mathcal{A}\ldots$ are the coefficients of the $SU(3)_C-SU(3)_C-U(1)_H$, $SU(2)_L-SU(2)_L-U(1)_H$, $U(1)_Y-U(1)_Y-U(1)_H$ anomalies. Moreover, $\mathcal{A}_{YHH}, \mathcal{A}_{HHH}$ correspond to $U(1)_Y-U(1)_H-U(1)_H$, $U(1)_H-U(1)_H-U(1)_H$ anomalies.

The final constraint comes from the vanishing of the mixed anomaly quadratic in the horizontal charges
\begin{align}\label{condi5}
\mathcal{A}_{YHH}=H^{2}_{u}-H^{2}_{d}+\sum_{i}\left[Q^{2}_{i}-L^{2}_{i}-2u^{2}_{i}+d^{2}_{i}+l^{2}_{i}\right]=0.
\end{align}

Given the above set of conditions, 13 out of 17 $H$ charges are constrained and can be expressed in terms of the remaining four free parameters that we choose to be $n_i$ $(i=1,2,3)$ and $x$. Where $x=H_d + Q_3 + d_3 = H_d + L_3 + l_3$ consistently with our parameterization of $\tan\beta=\theta^{x-3}$ such
that it ranges from 90 to 1 for $x$ running from 0 to 3 (see Ref.
~\cite{Mira:2000gg} for more details). The expressions for the standard model field $H$ charges are shown in Table~\ref{tab:char}. As can be seen from Table~\ref{tab:char}, the $H$ charges $n_i$ and $x$ act as free parameters and their possible values should be fixed by additional experimental constraints.
\begin{table}[t]
  \centering
  \begin{tabular}{rl}\hline\\
    $Q_3=$&$\displaystyle -\frac{-3 x (x+10)+(x+4) n_1+(x+7) n_2+(x+9)
        n_3-67}{15 (x+7)}$\\[0.5cm]
    $L_3=$&$\displaystyle \frac{2 (x+1) (3 x+22)-(2 x+23) n_1-2 (x+7)
      n_2+(13 x+97) n_3}{15 (x+7)}$\\[0.5cm]
    $L_2=$&$L_3+n_2-n_3$\\
    $L_1=$&$L_3+n_1-n_3$\\
    $H_u=$&$n_3-L_3$\\
    $H_d=$&$-1-H_u$\\
    $u_3=$&$-Q_3-H_u$\\
    $d_3=$&$-Q_3-H_d+x$\\
    $l_3=$&$-L_3-H_d+x$\\
    $Q_1=$&$3+Q_3$\\
    $Q_2=$&$2+Q_3$\\
    $u_1=$&$5+u_3$\\
    $u_2=$&$2+u_3$\\
    $d_1=$&$1+d_3$\\
    $d_2=$&  $d_3$\\
    $l_1=$&  $5-n_1+n_3+l_3$\\
    $l_2=$&  $2-n_2+n_3+l_3$\\\\\hline
  \end{tabular}
  \caption{Standard model fields $H$--charges in terms of 
    the bilinear $H$--charges $n_i$ and $x$}
  \label{tab:char}
\end{table}
With all these restrictions, there is only a possible set of charge
differences which is displayed in Table~\ref{tab:chargediff}. This
self-consistent solution includes the Guidice-Masiero mechanism to
solve the $\mu$ problem because $n_0=-1$, and therefore the $\mu$ term
is absent from the superpotential~\cite{Mira:2000gg}.
\begin{table}[t]
  \centering
  \begin{tabular}{cccccccc}\hline
    $Q_{13}$&$Q_{23}$&$d_{13}$&$d_{23}$&$u_{13}$&$u_{23}$&$\mathcal{L}_{13}$&$\mathcal{L}_{23}$\\\hline
    $3$&$2$&$1$&$0$&$5$&$2$&$5$&$2$\\\hline    
  \end{tabular}
  \caption{Standard model fields $H$ charge differences with $n_0=-1$ (from Ref. \cite{Mira:2000gg}). Here $\mathcal{L}_{i3}=L_{i3}+l_{i3}$}
  \label{tab:chargediff}
\end{table}

\section{Determination of the parameter of expansion $\theta$}\label{theta}
The vacuum expectation value of the flavon $\langle S \rangle$ is determined dynamically thanks to the anomalous nature of $U(1)_H$~\cite{Dreiner:2003hw}. We show explicitly that our $U(1)_H$--charges assignments can successfully lead to an expansion parameter given by $\theta=\langle S \rangle/M_{P}\approx 0.185-0.221$~\cite{Dreiner:2003yr,Irges:1998ax} as desired phenomenologically.
In the string-embedded FN framework the expansion parameter $\theta$ has its origin solely in the Dine-Seiberg-Wen-Witten mechanism, due to which the coefficient of Fayet-Iliopoulos (FI) is radiatively generated~\cite{Dreiner:2003hw, Cvetic:1998gv,Fayet:1974jb}
\begin{align}\label{c1}
\varepsilon_H=g_s^{2}\frac{\mathcal{A}_{GGH}}{192\pi^{2}}M_{\text{P}}^{2},
\end{align}
where $\mathcal{A}_{GGH}=Grav-Grav-U(1)_H$ is the gravitational anomaly, and $g_s$ being the string couplings constant.
The cancellation of the mixed chiral anomalies of $U(1)_H$ with the gauge group of the SM, itself and gravity demands, see Ref.~\cite{Maekawa:2001uk}
\begin{align}\label{c2}
\frac{\mathcal{A}_{CCH}}{k_C}=\frac{\mathcal{A}_{LLH}}{k_L}=\frac{\mathcal{A}_{YYH}}{k_Y}=\frac{\mathcal{A}_{HHH}}{3k_H}=\frac{\mathcal{A}_{GGH}}{24},
\end{align}
the $k$$\ldots$ are the affine or Kac-Moody levels of the corresponding symmetry~\cite{Dreiner:2003hw}.
Relying on the Green-Schwartz mechanism~\cite{Green:1984sg}, the term
\begin{align}\label{c3}
\mathcal{A}_{YHH}=0.
\end{align}
The factor of 3 in the fourth denominator in Eq.~(\ref{c2}) is of a combinatorial nature: one deals with a pure rather than mixed anomaly. 
In this convention one has:
\begin{align}\label{c4}
g_C^{2}k_C=g_L^{2}k_L=g_Y^{2}k_Y=g_H^{2}k_H=2g_s^{2},
\end{align}
$g_C$ being the $SU(3)_C$ couplings constant, $g_L$ being the $SU(2)_L$ couplings constant, $g_Y$ and $g_H$ are the $U(1)_Y$ and $U(1)_H$ couplings constant. For the factor of 2 in Eq.~(\ref{c4}) and a discussion of the mismatch between the conventions of GUT and string amplitudes see Ref.~\cite{Cvetic:1998gv}
($\varepsilon_H^{\text{tree level}}$ is zero in local supersymmetry, see Ref.~\cite{Barbieri:1982ac}). This gives
\begin{align}\label{c5}
\langle S \rangle=\sqrt{-\frac{\varepsilon_H}{S}},
\end{align}
supposing that no other fields break $U(1)_H$. With $S=-1$, we use the Eq.~(\ref{c2}) to eliminate $\mathcal{A}_{GGH}$ in favor of $\mathcal{A}_{CCH}$, 
\begin{align}\label{c6}
\frac{\mathcal{A}_{CCH}}{k_C}&=\frac{\mathcal{A}_{GGH}}{24} \nonumber \\
\mathcal{A}_{GGH}&=\frac{24}{k_C}\mathcal{A}_{CCH}.
\end{align}
Replacing the Eq.~(\ref{c6}) in the Eq.~(\ref{c1}),
\begin{align}\label{c7}
\varepsilon_H=g_s^{2}\frac{24}{192\pi^{2}k_C}\mathcal{A}_{CCH}M_{\text{P}}^{2},
\end{align}
check according to Ref.~\cite{Dreiner:2003hw}, we have
\begin{align}\label{c8}
\mathcal{A}_{CCH}=\frac{1}{2}\left[\sum_i(2Q_{i}+u_{i}+d_{i})\right].
\end{align}
Utilizing the charges shown in Table~\ref{tab:char}, we find that
\begin{align}\label{c9}
\mathcal{A}_{CCH}=\frac{3}{2}(7+x),
\end{align}
from the Eq.(\ref{c4}), we use
\begin{align}\label{c10}
g_C^{2}k_C&=2g_s^{2} \nonumber \\
g_s^{2}&=\frac{g_C^{2}k_C}{2}.
\end{align}
Replacing the Eq.(\ref{c9}) and the Eq.~(\ref{c10}) into the Eq.~(\ref{c7}), 
\begin{align}\label{c11}
\varepsilon_H&=\frac{24(7+x)3g_C^{2}k_C}{768\pi^{2}k_C}M_{\text{P}}^{2}\\ \nonumber
&=\frac{3(7+x)g_C^{2}}{32\pi^{2}}M_{\text{P}}^{2}.\\ \nonumber
\end{align}
Introducing the Eq.~(\ref{c11}) into the Eq.~(\ref{c5}), 
\begin{align}\label{c12}
\langle S \rangle=\frac{g_C}{4\pi \sqrt{2}}\sqrt{3(7+x)}M_{\text{P}},
\end{align}
and evaluating $g_C$ $\left[M_{\text{GUT}}=2.2 \times 10^{16}\text{GeV}\right]\approx 0.72$ and replacing in Eq.~(\ref{c12}), 
\begin{align}\label{c13}
\langle S \rangle=\frac{0.72}{4\pi \sqrt{2}}\sqrt{3(7+x)}M_{\text{P}},
\end{align}
the parameter $\theta$ in the Eq.~(\ref{1}) with $M_F=M_{\text{P}}$ and the Eq.~(\ref{c13}) is given by
\begin{align}\label{c14}
\theta=\frac{0.72}{4\pi \sqrt{2}}\sqrt{3(7+x)}.
\end{align}

We show in Table~\ref{tab:epsi} the variation of $\theta$ according to Eq.~(\ref{c14}) for different values of $x$.
\begin{table}[t]
  \centering
  \begin{tabular}{cccccc}\hline
    $\theta$& &$0.185$&$0.198$&$0.210$&$0.221$\\ \hline
    $x$& &$0$&$1$&$2$&$3$\\ \hline    
  \end{tabular}
  \caption{Variation of $\theta$ with the different values of $x$.}
  \label{tab:epsi}
\end{table}

 Finally, the reason for obtaining the condition $|H(f_i)|\lesssim10$ is to avoid an excessive fine--tuning in the Eq.(\ref{c8}).


\section{Relations for the charges of the model}\label{modelt}

By using Table~\ref{tab:char} is easy to check that
\begin{align}\label{ma}
H(\lambda_{ijk})=&\\
\begin{pmatrix}
  n_{\lambda_{121}} & n_{\lambda_{122}} & n_{\lambda_{123}} \\
  n_{\lambda_{131}} & n_{\lambda_{132}} & n_{\lambda_{133}} \\
  n_{\lambda_{231}} & n_{\lambda_{232}} & n_{\lambda_{233}} \\
\end{pmatrix}=&\begin{pmatrix}\nonumber
                x+n_{2}+6 & x+n_{1}+3 & x+n_{1}+n_{2}-n_{3}+1 \\
                x+n_{3}+6 & x+n_{1}-n_{2}+n_{3}+3 & x+n_{1}+1  \\
                x-n_{1}+n_{2}+n_{3}+6 & x+n_{3}+3 &  x+n_{2}+1  \\
              \end{pmatrix}.\nonumber
\end{align}
\begin{table}
  \centering
  \begin{tabular}{llll}\hline
    $i$ & $1$ & $2$ & $3$\\\hline
    $p_i$ & $3$ & $2$ & $2$\\
    $p_i'$ & $4$ & $1$ & $0$\\
    $p_i''$ & $3$ & $2$ & $2$\\\hline
  \end{tabular}
  \caption{Integer values required to obtain the horizontal charges of dimension-4 RPV operators.}
  \label{tab:pi}
\end{table}
This charges can be parameterized as:
\begin{align}\label{ca1}
H(\lambda_{ijk})=n_{i}+n_{j}-n_{k}+n_{0}+x+i-2k+p_i+p_j+p_k
\end{align}
the $p_i$ are given in the Table~\ref{tab:pi}. There are two possibilities for these charges,

\begin{enumerate}
\item $\lambda_{ijk}$ con $i=k$ or $j=k$.
We can write the Eq.~(\ref{ca1}) as
\begin{align}\label{ca2}
H(\lambda_{ijk})=n_{i(\text{or} j)}+n_{0}+x+i-2k+p_i+p_j+p_k,
\end{align}
the last five terms in the Eq.~(\ref{ca2}) can be seen as an integer function of the $i,j,k$ indices,
\begin{align}\label{ca3}
\mathcal{I}(i,j,k)=&i-2k+p_i+p_j+p_k\,,     &(i<&j)
\end{align}
and Eq.~(\ref{ca2}),
\begin{align}\label{ca4}
H(\lambda_{ijk})=n_{i(or j)}+n_{0}+[x+\mathcal{I}(i,j,k)].
\end{align}

\item $\lambda_{ijk}$ con $i\neq k$ or $j\neq k$
\begin{center}
\begin{align}\label{ca5}
H(\lambda_{ijk})=n_{i}+n_{j}-n_{k}+n_{0}+[x+\mathcal{I}(i,j,k)],
\end{align}
adding and subtracting $n_k$ in Eq.~(\ref{ca5})
\begin{align}\label{ca6}
H(\lambda_{ijk})=n_{i}+n_{j}+n_{k}-2n_{k}+n_{0}+[x+\mathcal{I}(i,j,k)],
\end{align}
the Eq.~(\ref{ca6}) can be written as
\begin{align}\label{ca7}
H(\lambda_{ijk})=n_{1}+n_{2}+n_{3}+n_{0}-2n_{k}+[x+\mathcal{I}(i,j,k)].
\end{align}
\end{center}
Simplifying
\begin{align}\label{ca8}
H(\lambda_{ijk})=\mathcal{N}-2n_{k}+[x+\mathcal{I}(i,j,k)],
\end{align}
where,
\begin{align}\label{ca9}
\mathcal{N}=\sum_{\alpha=0}^{3}n_{\alpha}=n_1+n_2+n_3+n_{0}.
\end{align}
\end{enumerate}
By using Eq.~(\ref{ca8}) and Eq.~(\ref{ca9}) we can reproduce any entry of the matrix in Eq.~(\ref{ma}), for example
\begin{align}\label{ca10}
H(\lambda_{123})=&n_{1}+n_{2}+n_{3}-1-2n_{3}+[x+1-6+3+4]\\
&=x+n_1+n_2-n_3+1.\nonumber
 \end{align}

In the same way the charges of the other couplings of Eq.~(\ref{po}) can be written as
\begin{align}\label{otros}
H(\lambda''_{ijk})=&\tfrac{1}{3}\mathcal{N}+\left[x+\mathcal{I}''(i,j,k)\right]&(j<&k) \\
H(\lambda'_{ijk})=&n_i+\left[x+\mathcal{I}'(i,j,k)\right], \nonumber   
\end{align}
where 
\begin{align}
  \label{eq:Is}
  \mathcal{I}''(i,j,k)=&-2i+p_i''+p_j''+p_k''&(j<&k)\nonumber\\
 \mathcal{I}'(i,j,k)=&\frac{1}{2}\left(j+k+p_j'+p_k'\right)-2\delta_{j3}.&&     \nonumber\\
\end{align}

To summarize, the $H$ charge of the $R$-parity breaking
couplings can be written as 
\begin{align}
\label{resum}
  H(\lambda''_{ijk})=&\tfrac{1}{3}\mathcal{N}+\left[x+\mathcal{I}''(i,j,k)\right]&(j<&k)\nonumber\\
  H(\lambda'_{ijk})=&n_i+\left[x+\mathcal{I}'(i,j,k)\right]                      &&     \nonumber\\
  H(\lambda_{ijk})=&n_i+n_j-n_k+n_0+\left[x+\mathcal{I}(i,j,k)\right]           &(i<&j)\nonumber\\
  =&
  \begin{cases}
    n_{i(\text{or $j$})}+n_0+\left[x+\mathcal{I}(i,j,k)\right]& \text{if $i= k$ (or $j= k$)}\\
    \mathcal{N}-2n_k+\left[x+\mathcal{I}(i,j,k)\right]& \text{if $i\ne k$ and $j\ne k$}\\
  \end{cases}.&&
\end{align}
From Eq.~(\ref{resum}) is straightforward to see the possible
scenarios that we can obtain in the context of an anomalous horizontal Abelian symmetry
with a single flavon as will be explained below:
\subsection{Getting the MSSM}\label{u-mssm}

It was shown in the Sec.~\ref{gag} that the conservation of $R$--parity in the MSSM $+\widehat{N}$,
results as a consequence of the proper choice of the horizontal $U(1)_{H}$--charges 
and based on this it is possible to obtain a proper mass texture for the neutrinos. To recover the MSSM, we need that the bilinear terms ($\mu_{i})$ and the trilinear terms ($\lambda_{ijk}, \lambda'_{ijk}, \lambda''_{ijk}$) which violate $R$--parity are prohibited. To prohibit the bilinear terms along with $\lambda'_{ijk}$ and $\lambda_{ijk}$ with repeated indices. If $\mathcal{N}$ in Eq.~(\ref{resum}) is a rational number, we obtain $\mathcal{N}/3$ is fractional, then the term $\lambda''_{ijk}$ is also forbidden. From Eq.~(\ref{31}), $n_i$ must be half--integer and therefore $\mathcal{N}$ is fractional in order to have $\lambda_{ijk}$ fully prohibited. These conditions are fulfilled if we choose for example $n_1=-3/2$, $n_2=-5/2$, $n_3=-5/2$ and $x=2$. With this choice, we obtain the charges shown in the Table~\ref{mssm-n} which are in agreement with Ref.~\cite{Dreiner:2003yr}.
By using these charges and the Eqs.~(\ref{resum}) are forbidden the trilinear terms in the Eq.~(\ref{po}). For Example:

\begin{align}\label{ca11}
    H(\lambda_{ijk})=&-3/2-5/2-5/2-1+5+[x+\mathcal{I}(ijk)]\\
=&-5/2+[x+\mathcal{I}(ijk)],\nonumber\\
H(\lambda'_{ijk})=&-5/2+\left[x+\mathcal{I}'(i,j,k)\right],\nonumber\\
H(\lambda''_{ijk})=&\tfrac{1}{3}(-3/2-5/2-5/2-1)+\left[x+\mathcal{I}''(i,j,k)\right]\nonumber\\
=&-15/6+\left[x+\mathcal{I}''(i,j,k)\right].\nonumber
\end{align}
\vspace{6.0cm}

\begin{table}[t]
  \centering
  \begin{tabular}{cccccc}\hline
                Generation $i$&$Q_{i}$& $d_{i}$ & $u_{i}$ & $L_{i}$ & $l_{i}$ \\\hline 
    1 & 67/15 & 13/30 & 169/30 & 3/5 & 53/10 \\ \hline 
    2 & 52/15 & -17/30 & 79/30 & -2/5& 33/10   \\ \hline 
    3 & 22/15 & -17/30 & 19/30 &-2/5 &13/10 \\ \hline 
      \end{tabular}
\end{table}
\begin{table}[t]
  \centering
  \begin{tabular}{cccc}\hline
              $H_{u}$&$H_{d}$& $N_{1}$ & $N_{2}$  \\\hline 
    -21/10 & 11/10 & 5/2 & 5/2  \\ \hline 
      \end{tabular}
      \caption{The set complete of $H$--charges for $x=2$. It gives a texture for the neutrinos sector and it reproduce the MSSM $+\widehat{N}$.}
  \label{mssm-n}
\end{table}

Now, the superpotential which introduces the interaction terms for both Dirac and Majorana neutrinos is
\begin{align}\label{majo}
    \mathcal{W}^{\nu}&=M_{\text{P}}\frac{M_{ij}}{2}\theta^{N_{i} +
    N_{j}}N^{i}N^{j} +  Y^{\nu}_{ij}\theta^{L^{i} + H^{u}
    + N^{j}}L^{i}H^{u}N^{j}\\ \nonumber
    &+ \frac{\psi_{ij}}{2M_{\text{P}}}\theta^{L^{i} +
    H^{u} + L^{j}+ H^{u}}
    L^{i}H^{u}L^{j}H^{u},
\end{align}
the various terms of mass for neutrinos in this superpotential are, in order of appearance, as follows:
$\mathbf{M_{RR}^{Maj}}$, $\mathbf{M_{LR}^{Dirac}}$ and
$\mathbf{M_{LL}^{Maj}}$.

Note that $n_i$ are less than zero. Then the term $\mathbf{M_{LL}^{Maj}}$, with $H$ charges $(n_i +n_j )$,  is suppressed by a factor of $m_{3/2}/M_{\text{P}}$, and therefore it can be disregarded in the Eq.~(\ref{majo}), the rest are just the terms to build the seesaw mechanism:
\begin{align}\label{see}
     M^{\nu}_{ij}=-\theta^{L^{i} + L^{j} +
    2H^{u}}\left(
    \sum_{k,l}Y^{\nu}_{ik}M^{-1}_{kl}Y^{\nu}_{jl}\right).
\end{align}
The term $\theta^{L^{i}+L^{j}+2H^{u}}$ is greater than one, and so the $m_{\nu}$ can be improved by the parameter $\theta$, enhancing the consistency with phenomenology. According to the analysis, the set of $H$--charges given in 
Table~\ref{mssm-n}, give a proper texture matrix for neutrinos
\begin{align}\label{mat}
    \mathbf{M}^{\nu}\sim \theta^{-5}\langle H^{u} \rangle^{2}/M_{P}\begin{pmatrix}
      \theta^{2} & \theta & \theta \\
      \theta & 1 & 1 \\
      \theta & 1 & 1 \\
    \end{pmatrix}.
\end{align}
Note that the seesaw scale is obtained from the single scale of the model $M_{P}$. In this thesis, this mechanism will be generalized to the case of $SU(5)$ SUSY~\cite{Chen:2008tc,Jack:2003pb,Babu:2002tx,Ling:2002nj,Bando:2003iz}.
\subsection{Getting bilinear $R$-parity violation}
The bilinear $R$--Parity violating models are characterized by two properties~\cite{Hirsch:2000ef,Diaz:2003as}: first, the usual MSSM superpotential is enlarged according to
\begin{align}\label{bili}
W_{\text{BRPV}}=W_{MSSM} + \varepsilon^{ab}\mu^{i}\widehat{L}^{i}_{a}\widehat{H}^{u}_{b},
\end{align}
where there are 3 new superpotential parameters $(\mu^{i})$, one for each fermion generation. The second modification is the addition of extra soft term

\begin{align}\label{biliso}
V_{\text{soft}}=V_{MSSM} - \varepsilon^{ab}B^{i}\mu^{i}\tilde{L}^{i}_{a}H^{u}_{b},
\end{align}
that depends on three soft mass parameters $B^{i}$. For the sake of simplicity it is considered the $R$--conserving soft terms as in minimal supergravity (mSUGRA). Notice that the presence of the new soft interactions prevents the new bilinear terms in Eq.~(\ref{bili}) to be rotated away~\cite{Diaz:1997xc}.
The new bilinear terms break explicitly $R$--parity as well as lepton number. The bilinear $R$--parity violating models predicts correlations between observables in accelerators and neutrino physics~\cite{Porod:2000hv,deCampos:2012pf,DeCampos:2010yu} and they are sought at the LHC~\cite{ATLAS:2011ad}.

In our model, bilinear $R$-parity violation is obtained if we choose $\mathcal{N}/3$ fractional in Eq.~(\ref{resum}). We achieve this if $\mathcal{N}$ is rational, then $\lambda''_{ijk}$ is prohibited. If for each $n_i\lesssim-7$ in Eq.~(\ref{resum}) we get that the terms $\lambda'_{ijk}$ and $\lambda_{ijk}$ remain suppressed by a factor of the order of $m_{3/2}/M_{P}$.
For example, following the Eq.~(\ref{ca9}) and using the charges in the Tables~\ref{bilin2} and ~\ref{bilin} which are derived through the four free parameters $n_1=n_2=-7$, $n_3=-8$, $x=0$~\cite{Dreiner:2006xw} and $n_1=n_2=n_3=-8$, $x=1$~\cite{Mira:2000gg} . We obtain,
\vspace{2.0cm}
\begin{table}[t]
  \centering
  \begin{tabular}{cccccc}\hline
                Generation $i$&$Q_{i}$& $d_{i}$ & $u_{i}$ & $L_{i}$ & $l_{i}$ \\\hline 
    1 & 467/105 & -97/35& 722/105 & -386/105 & 667/105 \\ \hline 
    2 & 467/105 & -167/35 & 302/105 & -386/105& 352/105   \\ \hline 
    3 & 257/105 & -167/35 & 92/105 &-491/105 &247/105 \\ \hline 
      \end{tabular}
\end{table}
\begin{table}[t]
  \centering
  \begin{tabular}{cc}\hline
              $H_{u}$&$H_{d}$  \\\hline 
    -349/105 & 244/105   \\ \hline 
      \end{tabular}
      \caption{ Charges that enable bilinear $R$-parity violation with $x=0$ and $n_1=n_2=-7$,$n_3=-8$, according to the Ref.~\cite{Dreiner:2006xw}}
  \label{bilin2}
\end{table}

\begin{table}[t]
  \centering
  \begin{tabular}{cccccc}\hline
                Generation $i$&$Q_{i}$& $d_{i}$ & $u_{i}$ & $L_{i}$ & $l_{i}$ \\\hline 
    1 & 161/30 & -18/5 & 103/15 & -113/30 & 98/15 \\ \hline 
    2 & 131/30 & -23/5 & 58/15 & -113/30& 53/15   \\ \hline 
    3 & 71/30 & -23/5 & 28/15 &-113/30 &23/15 \\ \hline 
      \end{tabular}
\end{table}
\begin{table}[t]
  \centering
  \begin{tabular}{cc}\hline
              $H_{u}$&$H_{d}$  \\\hline 
    -127/30 & 97/30   \\ \hline 
      \end{tabular}
      \caption{Charges that enable bilinear $R$-parity violation with $x=1$ and $n_1=n_2=n_3=-8$, according to the Ref.~\cite{Mira:2000gg}}
  \label{bilin}
\end{table}
\vspace{-2.5cm}
\begin{align}\label{ca12}
\frac{\mathcal{N}}{3}=\frac{-8-8-8-1}{3}=\frac{-25}{3},\\
\frac{\mathcal{N}}{3}=\frac{-7-7-8-1}{3}=\frac{-23}{3}.\nonumber
\end{align}

This condition ensures that, for some bilinear charges $n_{i}\lesssim-7$, the $L$ violating trilinear terms in the Eq.~(\ref{po}) are very suppressed, while the $B$ violating are forbidden.
 
\subsection{Getting a $R$-parity breaking model with $L$ violation}
Following the Ref.~\cite{Sierra:2009zq}, and assuming a decaying neutralino as dark matter candidate, it is studied the neutralino decays in the context of the minimal $R$--parity violating models with only lepton number violating $\lambda$. The lifetime of a mainly gaugino neutralino decaying through a trilinear $R$--parity breaking coupling $\lambda$ is approximately given by (see Ref.~\cite{Baltz:1997ar})
\begin{align}\label{ca13}
\tau_{\chi}=\left(\frac{M_s}{2 \times 10^{4}\text{GeV}}\right)^{4}\left(\frac{10^{-23}}{\lambda}\right)^{2}\left(\frac{2 \times 10^{3}\text{GeV}}{m_{\chi}}\right)^{5}10^{26}\;\text{sec}.
\end{align}
According to this expression the viability of neutralino decaying DM will depend, for a few TeV neutralino mass, on the slepton mass spectrum and the size of the corresponding $\lambda$ coupling that will be determined by the choices $n_{\lambda}<0$. Due to the strong suppression induced by the factor $m_{3/2}/M_P$. A coupling as small as $10^{-23}$ is possible if $n_{\lambda}=-10$ and accordingly even with a not so heavy slepton the constraint $\tau_{\chi}\geq 10^{26}\ \text{sec}$ can be satisfied. To get only the couplings $\lambda$, we need that $\mathcal{N}$ in Eq.~(\ref{resum}) be a rational number, with this condition we prohibit $\lambda''_{ijk}$. Now if the bilinear $n_{i}$ are not a half-integer fractional, the terms $\lambda'_{ijk}$ and the $\lambda_{ijk}$, with $\text{ $i= k$ (or $j= k$)}$, are also prohibited. However, the terms $\lambda_{ijk}$ with \text{if $i\ne k$ and $j\ne k$} may be allowed if $\mathcal{N}-2n_k$ is an integer number. In such a case, the
decays of the LSP are leptophilic~\cite{Sierra:2009zq}. A set of bilinear charges that satisfy this condition are shown in the Table~\ref{rvl}. 
As an example, if we use the Eqs.~(\ref{resum}) and Table~\ref{rvl}, the only $R$--parity violating coupling at all scales is
\begin{table}[t]
  \centering
  \begin{tabular}{lccccc}\hline
    & $x$& $n_1$   &$n_2$      &$n_3$     & $|f_i|$\\\hline
    $\lambda_{231}$& $1$& $7/3$ & $-19/3$  & $-25/3$&  $<7$ \\\hline 
    $\lambda_{123}$& $1$& $-10/3$ & $-19/3$  & $7/3$&  $<6$ \\\hline 
    $\lambda_{132}$& $1$& $-5/3$ & $17/3$  & $-20/3$&  $<7$ \\\hline 
  \end{tabular}
  \caption{Set of bilinear $H$--charges consistent with the trilinear 
    $H$--charge choice $n_\lambda=-10$.}
  \label{rvl}
\end{table}
\begin{align}\label{ca14}
    H(\lambda_{231})&=n_2+n_3+n_1-1-2n_{1}+\left[1+i-2k+p_i+p_j+p_k\right]\nonumber \\
&=-19/3-25/3+7/3-1-14/3+\left[1+2-2+2+2+3\right]\nonumber \\
&=-10,
\end{align}

while for example
\begin{align}\label{ca15}
    H(\lambda_{123})&=n_2+n_3+n_1-1-2n_{1}+\left[1+i-2k+p_i+p_j+p_k\right]\nonumber \\
&=-19/3-25/3+7/3-1+50/3+\left[1+1-6+3+2+2\right]\nonumber \\
&=19/3,
\end{align}

is a fractional number, along as for example
 \begin{align}\label{ca16}
    H(\lambda_{123}^{'})&=n_1+\left[1+\frac{1}{2}(j+k+p_{j}^{'}+p_{k}^{'}\right]\nonumber \\
&=7/3+\left[1+\frac{1}{2}(5+1)\right]\nonumber \\
&=19/3,
\end{align}
and
\begin{align}\label{ca17}
    H(\lambda_{123}^{''})&=\frac{1}{3}(7/3-19/3-25/3-1)+\left[1-2+3+2+2\right]\nonumber \\
&=7/3+\frac{1}{2}\left[5+1\right]\nonumber \\
&=14/9.
\end{align}

\subsection{Getting Majorana neutrinos with two flavons}

Here is also possible to have Majorana neutrinos if in addition to the right handed neutrinos we include in the model a second and third flavon, $\psi$, $\phi$, with a vacuum expectation value approximately equal to $\theta$. The horizontal charges of these fields are fixed by new invariant diagrams from Dirac and Majorana mass terms.
In this way, the $H$--charges $\psi$ and $\phi$ must be such that it does not get coupled to $L$ violating operators. Therefore, the respective overall $H$--charge of the $L$ violating operator would be fractional and therefore forbidden.
The introduction of two flavons field could spoil the proton stability since $H$--invariant terms can be obtained by coupling a large number of $\psi$ and $\phi$ flavons to dangerous operators. In our case, for the charges shown in the Table~\ref{mao}, all the dangerous operators that are coupled to new fields produce a overall fractional charge. 
Then by adding a second and third flavon field with fractional charge, it is possible to build a Majorana neutrino mass matrix. In both cases an anarchical matrix is obtained, see Refs.~\cite{Hall:1999sn,Haba:2000be,deGouvea:2003xe,deGouvea:2012ac}, which is supported by the recent results of a large value for $\theta_{13}$.
\begin{align}\label{anarq}
    \mathbf{M}^{\nu}\sim \theta^{-5}\langle H^{u} \rangle^{2}/M_{P}\begin{pmatrix}
      1 & 1 & 1 \\
      1& 1 & 1 \\
      1 & 1 & 1 \\
    \end{pmatrix}.
\end{align}
\begin{table}[t]
  \centering
  \begin{tabular}{ccccccccc}\hline
                & $n_{\lambda}$& $x$ & $n_1$ & $n_2$ & $n_3$ & $\psi$ & $\phi$ & $N_{i}$ \\\hline 
    $\lambda_{132}$ & -12 & 1 & -9973/1399 & 2438/1399 & -9973/1399 & -13270/1399 & -859/1399& 10832/1399\\ \hline 
    $\lambda_{132}$ & -12 & 2 & -9137/1213 & 2347/1213& -9137/1213 & -11972/1213 & -488/1213&9625/1213  \\ \hline 
    $\lambda_{231}$ & 13 & 3 & 4318/3907 & 9973/3907 &9973/3907 & -32078/3907&-37733/3907 &27760/3907  \\ \hline 
    $\lambda_{231}$ & 2 & 1 & -411/3907 & -9973/3907 &-9973/3907 & -22620/3907&-13058/3907  &23031/3907 \\\hline 
      \end{tabular}
      \caption{Examples set of $H$--charges which allow us having Majorana neutrinos. Where $N_i$ are the right-handed neutrinos.}
  \label{mao}
\end{table}

\subsection{Model with violation of baryon number}

In the Chapter~\ref{cap2}, we consider a supersymmetric standard model extended with an anomalous horizontal symmetry $U(1)_{H}$ of a single flavon. A self-consistent framework with baryon-number violation is achieved along with a proper suppression for lepton-number violating dimension-five operators, so that the proton can be sufficiently stable. With the introduction of right-handed neutrinos both Dirac an Majorana masses can be accommodated within this model.
In order to obtain a model with baryonic number violation we need that $\mathcal{N}$ in Eq.~(\ref{resum}) be multiple of 3. This condition ensures that the couplings $\lambda^{''}$ are generated. Choosing the bilinear terms $n_i$ fractional but not a half--integer, we guarantee that the $\lambda^{'}$ and $\lambda$ remain prohibited. For example, by choosing $n_1=n_2=n_3=13/3$ and $x=1$ we can see that only the $\lambda^{''}$ are generated. Using the Eq.~(\ref{resum}), we have for example

\begin{align}\label{ejj}
H(\lambda''_{323})=&\tfrac{1}{3}(13/3+13/3+13/3-1)+\left[1+0)\right]\\
=&5.\nonumber
\end{align}
In the same way all $\lambda^{''}$ are obtained.

We can check for example that

\begin{align}\label{ejj2}
H(\lambda'_{123})=&13/3+\left[1+3\right]\\
=&25/3, \nonumber
\end{align}

and
\begin{align}\label{ejj3}
H(\lambda_{231})&=(13/3+13/3+13/3-1)-2(13/3)+\left[1+7\right]\\
=&34/3,\nonumber
\end{align}
are fractional.

\section{$SU(5)\times U(1)_{H}$ with several flavons}\label{unif}

Differently from the SM case~\cite{Dreiner:2003yr}, in $SU(5)$ GUTs it
is rather difficult to implement the horizontal symmetries,
because there is less freedom in choosing the $H$--charges (see for
example~\cite{Chen:2008tc}). However, if we allow for several flavons that break the
horizontal symmetry, and they are assigned to the adjoint representation of
$SU(5)$~\cite{Aristizabal:2003zn,Duque:2008ah,Wang:2011ub}, charges
that were forbidden in the singlet flavon case become allowed, under
the assumption that certain representations for the FN~\cite{Froggatt:1978nt} messengers fields do not exist.  In
contrast to the non-unified $SU(3)_{C}\times SU(2)_{L}\times U(1)_{Y}\times
U(1)_H$ model, where the singlet nature of the flavons is mandatory,
in $SU(5)\times U(1)_H$ assigning the flavons to the adjoint has the
additional bonus that non-trivial group theoretical coefficients
concur to determine the coefficients of the effective
operators~\cite{Aristizabal:2003zn,Duque:2008ah,Wang:2011ub}. In this
case, under the additional assumption that at the fundamental level
all the Yukawa couplings obey to some principle of
universality~\cite{Duque:2008ah}, the order one coefficients that
determine quantitatively the structure of the mass matrices become
calculable. In this thesis we generalize the mechanism of obtain $R$--parity from an horizontal symmetry described in Sec.~\ref{u-mssm} to the context of $SU(5)$ SUSY $+ \widehat{N}$.

\chapter{Baryonic violation of $R$-parity from anomalous $U(1)_H$}\label{cap2}
\label{intro2}
Supersymmetric scenarios with $R$-parity conservation are becoming very constrained due to the lack of missing energy signals associated to heavy neutral particles, thus motivating scenarios with $R$-parity violation. In view of this, we consider a supersymmetric model with $R$-parity violation and extended by an anomalous horizontal $U(1)_{H}$ symmetry. A self-consistent framework with baryon-number violation is achieved along with a proper suppression for lepton-number violating dimension-five operators, so that the proton can be sufficiently stable. With the introduction of right-handed neutrinos both Dirac and Majorana masses can be accommodated within this model. The implications for collider physics are discussed.
\section{Horizontal model with Baryon-number violation}
\label{sec:BNVmodel}
In the simplest scenario, the $U(1)_H$ symmetry is spontaneously
broken at one scale close to Planck mass, $M_P$, by the vacuum expectation
value of a SM singlet scalar, the flavon field $S$, with $H$ charge~$-1$, which allows us to define the expansion parameter $\theta=\langle S
\rangle/M_P\approx0.22$ (see Sec.~\ref{theta}). 
The fermion masses and mixings are determined by factors of the type
$\theta^n$,for which $n$ is fixed by the horizontal charges of the fields
involved.
In supersymmetric scenarios, the order of magnitude of the $R$-parity
violating couplings can also be fixed by the FN mechanism
\cite{Mira:2000gg, Dreiner:2003hw, Dreiner:2003yr, Choi:1998wc,Choi:1996se, Binetruy:1996xk, Joshipura:2000sn, Ellis:1998rj, BenHamo:1994bq}.

In what follows we will constrain the $H$ charge to satisfy the
condition $|H(f_i)| \lesssim 10$ which leads to a consistent prediction of
the size of the suppression factor $\theta$ in the context of string
theories~\cite{Dreiner:2003hw,Choi:1996se}(see discussion in Sec.~\ref{theta}).

From Eq.~\eqref{resum} is straightforward to see the possible
scenarios in the context of an anomalous horizontal Abelian symmetry
with a single flavon, reviewed in the introduction.
The MSSM is obtained when $\mathcal{N}/3$, each individual $n_i$  
and $\mathcal{N}-2n_k$ are fractional~\cite{Dreiner:2003yr,Dreiner:2007vp}.
Bilinear $R$-parity violation\footnote{See, for example, Ref. \cite{Diaz:2003as} and references therein} is obtained when $\mathcal{N}/3$
is fractional and each $n_i$ is a negative integer ~\cite{Dreiner:2006xw, Mira:2000gg}.
Another self-consistent $R$-parity breaking model with $L$ violation can
be obtained if $\mathcal{N}/3$ and  each individual $n_i$ are fractional,
but some of the  $\mathcal{N}-2n_k$ are integers. In such a case the
decays of the LSP are leptophilic~\cite{Sierra:2009zq}.( For these developments see Sec.~\ref{modelt}).

In this thesis we want to explore the last self-consistent possibility,
consisting in the $R$-parity breaking model with $B$ violation.  It is
clear from Eq.~\eqref{resum} that if $\mathcal{N}$ is an integer and
multiple of 3, and each $n_i$ is fractional but not half-integer,
then only the 9 $\lambda_{ijk}''$ are generated.  The specific horizontal charges are
\begin{align}
\label{eq:Hlpp}
 H \begin{pmatrix}
    \lambda_{112}''&\lambda_{212}''&\lambda_{312}''\\
    \lambda_{113}''&\lambda_{213}''&\lambda_{313}''\\
    \lambda_{123}''&\lambda_{223}''&\lambda_{323}''\\
  \end{pmatrix}=
 \begin{pmatrix}
   6 & 3 & 1\\
   6 & 3 & 1\\
    5 & 2 & 0\\
  \end{pmatrix} +n_{\lambda''}\mathbf{1_{3}},
\end{align}
where  $\mathbf{1_3}$ is a $3\times 3$ matrix filled with ones, and  $n_{\lambda''}$  is defined by
\begin{align}\label{nlpp}
  n_{\lambda''}=x+ \frac{1}{3}\mathcal{N}\,.
\end{align}

For positive $n_{\lambda''}$ values, the third-generation couplings dominate with fixed ratios between them:
\begin{align}
\label{eq:Hierlpp}
 \begin{pmatrix}
    \lambda_{112}''&\lambda_{212}''&\lambda_{312}''\\
    \lambda_{113}''&\lambda_{213}''&\lambda_{313}''\\
    \lambda_{123}''&\lambda_{223}''&\lambda_{323}''\\
  \end{pmatrix}\approx&
 \theta^{n_{\lambda''}}\begin{pmatrix}
   \theta^6 & \theta^3 & \theta\\
   \theta^6 & \theta^3 & \theta\\
   \theta^5 & \theta^2 & 1\\
  \end{pmatrix} & n_{\lambda''}\ge 0\,.
\end{align}
For negative values some of the couplings start to be forbidden in the superpotential by
holomorphy, and for $n_{\lambda''}<-6$ all of them must be generated
from the K\"ahler potential with additional Planck mass suppression,
so that the LSP may be a decaying dark matter candidate as in the
case of $L$ violation studied in Ref.~\cite{Sierra:2009zq}.
We will not pursue this possibility in this work because in that case
the phenomenology at colliders should be the same as that in the MSSM.

Below the allowed range for $n_{\lambda''}$ and their
consequences at present and future colliders will be checked.

\subsection{Constraints from $\Delta B\neq0$ processes}

\begin{figure}
  \centering
  \includegraphics[scale=0.6]{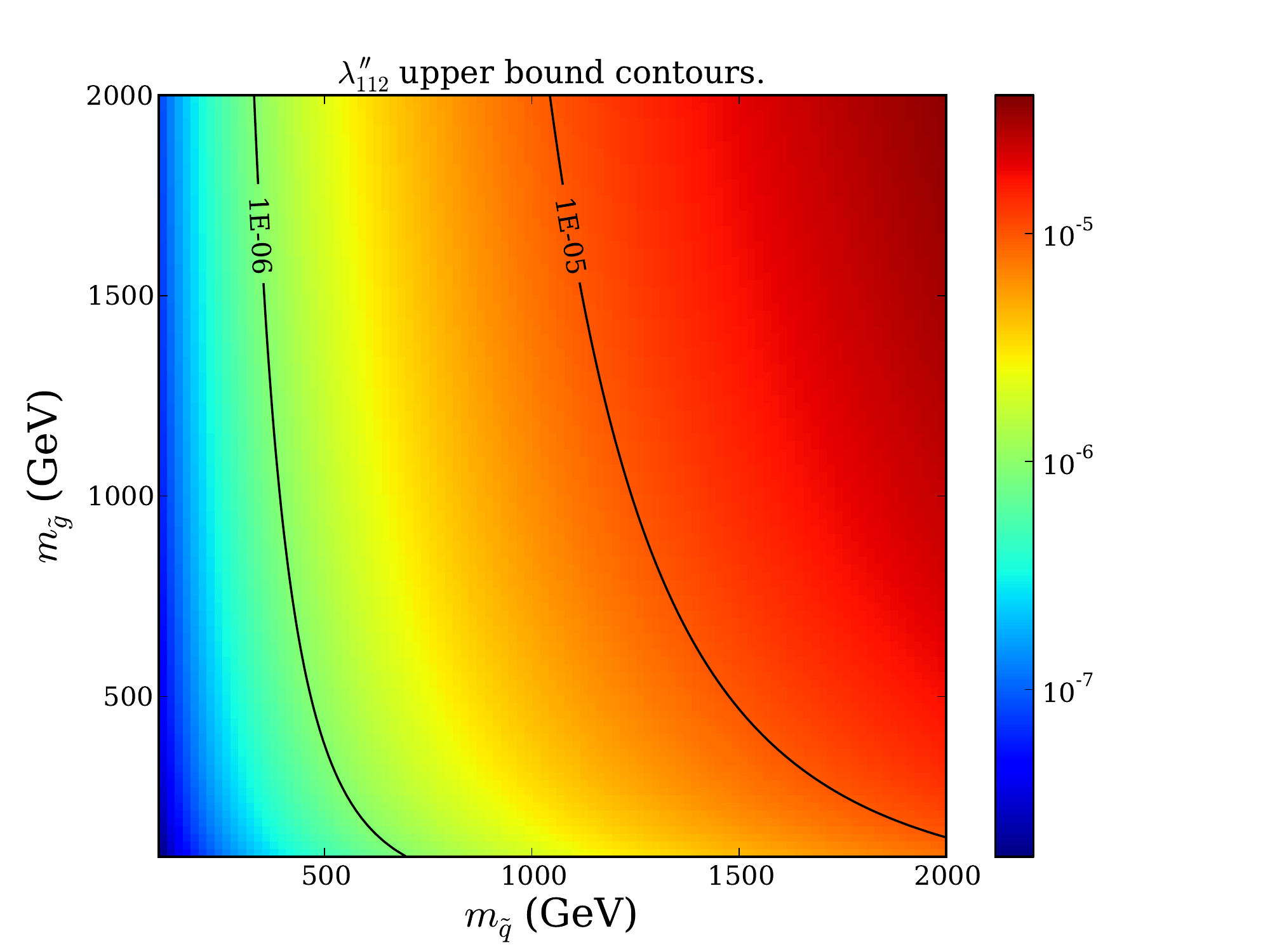}
  \caption{$\lambda_{112}''$ constraint as a function of squark and gluinos mass, for $\tilde{\Lambda}=75\ \text{GeV}$ and $\tau_{NN\to KK}=1.7\times10^{32}\, \text{yr}$}
  \label{fig:constraints_contour}
\end{figure}

Several experimental constraints are found on $B$ violating
couplings both for individual and quadratic products of couplings~\cite{Barbier:2004ez}. For individual couplings, the stronger constraints are for  $\lambda_{11k}$. Because in our model the predicted order of magnitude for the coupling $\lambda_{113}''$ is the same as that for $\lambda_{112}''$, the most restrictive constraint is that obtained for the later and comes from 
the dinucleon $NN\to KK$ width, which according to Refs.~\cite{Goity:1994dq,Csaki:2011ge} is
\begin{align}
  \Gamma\sim \rho_N\frac{128\pi\alpha_s^2|\lambda''_{112}|^4(\tilde{\Lambda})^{10}}{m_{N}^2m_{\tilde{g}}^2m_{\tilde{q}}^8}\,,
\end{align}
where $\rho_N\approx 0.25\ \text{fm}^{-3}$ is the nucleon density,
$m_N\approx m_p$ is the nucleon mass, and $\alpha_s\approx0.12$ is the
strong coupling.
Note that this kind of matter instability requires only $B$ violation and is suppressed by the tenth power of  $\tilde{\Lambda}$, which parametrizes the hadron and nuclear effects. 
For this quantity,  order of magnitude variation is expected around of the $\Lambda_{\text{QCD}}$ scale of $200\ \text{MeV}$. 
However, $\tilde{\Lambda}$ is roughly expected to be smaller than $\Lambda_{\text{QCD}}$ because of the repulsion effects inside the nucleus~\cite{Csaki:2011ge}.  
From general experimental searches of matter instability~\cite{Berger:1991fa}, lower bounds similar to the proton lifetime should be used for this specific dinucleon channel~\cite{Goity:1994dq}, and therefore additional suppression from $\lambda_{112}$ could be required. In fact, the first lower bound on dinucleon decay to kaons has been recently obtained from Super-Kamiokande data~\cite{Litos:2010}
\begin{align*}
  \tau_{NN\to KK} =\frac{1}{\Gamma}>1.7\times10^{32}\ \text{yr}\,.
\end{align*}
From this value, we can obtain  a constraint for the $B$ violating coupling: 
\begin{align}
  |\lambda''_{112}|\lesssim3.2\times 10^{-7}
\left(\frac{1.7\times10^{32}\, \text{yr}}{\tau_{NN\to KK}}\right)^{1/4}
\left(\frac{m_{\tilde{g}}}{300\, \text{GeV}}\right)^{1/2}
\left(\frac{m_{\tilde{q}}}{300\, \text{GeV}}\right)^{2}
\left(\frac{75\ \text{MeV}}{\tilde{\Lambda}}\right)^{5/2}\,,
\end{align}
where a conservative value for $\tilde{\Lambda}$, as in \cite{Barbier:2004ez}, has been used. Large values of $\tilde\Lambda$ give rise to even smaller upper bounds for $|\lambda_{112}|$.  In Fig.~\ref{fig:constraints_contour}, we illustrate the effect of varying gluino and squark masses.  We can see that the constraint still holds strong for large values of the relevant supersymmetric masses, especially for low-mass gluinos. 

\begin{figure}
  \centering
  \includegraphics[scale=0.7]{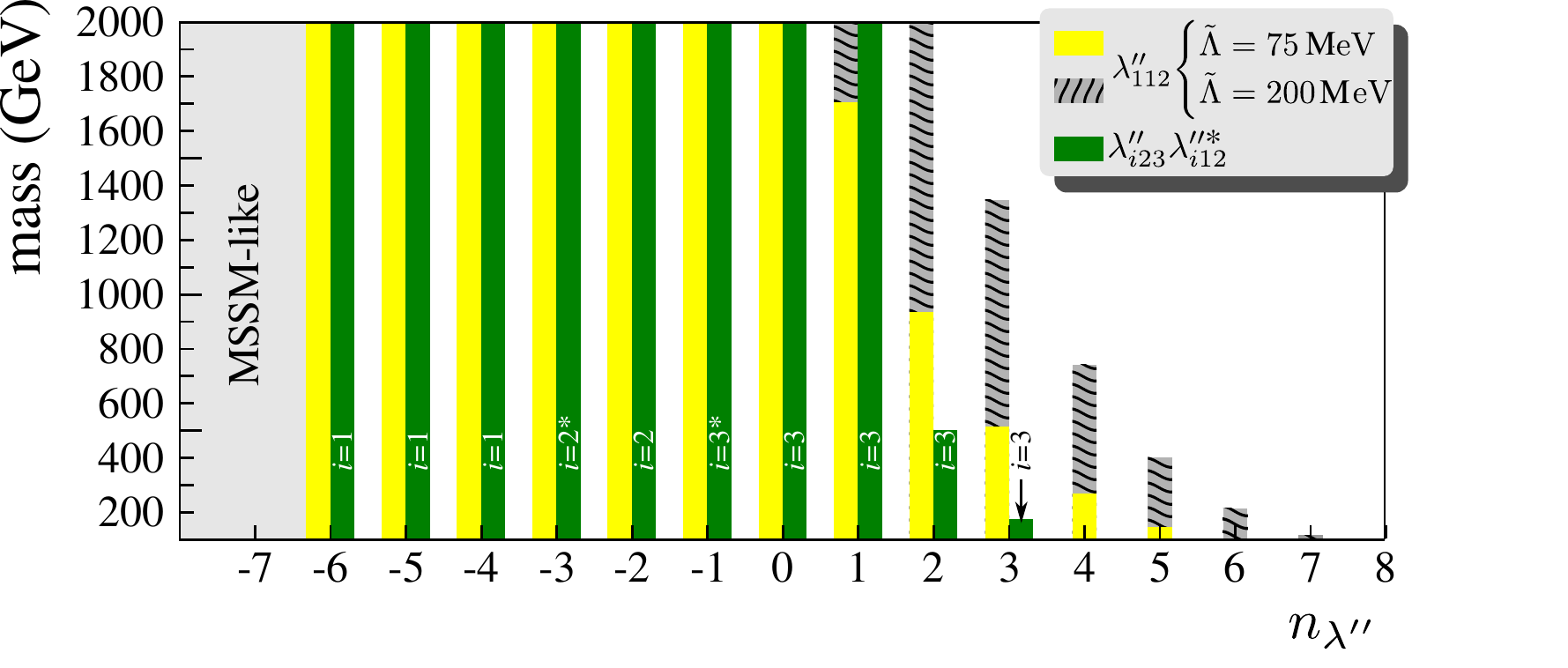}
  \caption{The yellow bands (green bands) display the excluded range for 
    $\tilde{m}$ ($\tilde{u}_{iR}$), as a function of the possible $n_{\lambda''}$ 
    solutions from the constraints in 
    $|\lambda_{112}''|$ ($|\lambda_{i23}''\lambda_{i12}^{\prime\prime *}|$). 
    The gray dashed bands show the effect of increasing
    $\tilde{\Lambda}$ in the $|\lambda_{112}''|$ constraint. 
    The affected $\tilde{u}_{iR}$ is indicated for each value of
    $n_{\lambda''}$. 
    For $n_{\lambda''}<-6$, the phenomenology at
    colliders is expected to the same as in the MSSM. }
  \label{fig:constraints}
\end{figure}

For $\tilde{m}=m_{\tilde{g}}=m_{\tilde{q}}$, we can obtain the lower bound
\begin{align}
  \label{eq:tildem}
 \tilde{m}  \gtrsim& (279\ \text{GeV})\theta^{(-8+2n_{\lambda''})/5}
   \left(\frac{\tau_{NN\to KK}}{1.7\times10^{32}\, \text{years}}\right)^{1/10}
   \left(\frac{\tilde{\Lambda}}{75\ \text{MeV}}\right),&n_{\lambda''}\ge& -6\,.
\end{align}
The excluded supersymmetric masses as function of $n_{\lambda''}$ are illustrated with the yellow (light-gray) bands in Fig.~\ref{fig:constraints}. The important restrictions appear for negative powers of $\theta$ in Eq.~\eqref{eq:tildem}, corresponding to $n_{\lambda''}\le 4$. If $\tilde\Lambda$ is increased to $150\ \text{MeV}$,  stronger restrictions are obtained, as illustrated in the dashed bands of Fig.~\ref{fig:constraints}. We can see that for the full range of equal gluino and squark  masses displayed in figure~\ref{fig:constraints}, the constraint is strong enough to forbid all the negative solutions of $n_{\lambda''}$ and also some of the positive solutions depending of the chosen $\tilde{\Lambda}$ value. 

It is also possible to exclude the negative solutions if we use the available quadratic coupling product bounds. For our model the most important constraint is obtained from the penguin decays $B\to \phi\pi$~\cite{Barbier:2004ez, BarShalom:2002sv}. Updating the limit with the last result from BABAR~\cite{Aubert:2006nn}\footnote{The limit from Belle is $\operatorname{Br}(B^+\to \phi\pi^+)<3.3\times 10^{-7}$~\cite{Kim:2012gt}.} to $\operatorname{Br}(B^+\to \phi\pi^+)<2.4\times10^{-7}$, we obtain from Fig. 3 of Ref.~\cite{BarShalom:2002sv}
\begin{align}
  |\lambda_{i23}''\lambda_{i12}^{\prime\prime *}|<2\times 10^{-5}\left(
    \frac{m_{\tilde{u}_{i R}}}{100\text{ GeV}}
  \right)^2\,.
\end{align}
The excluded right-handed up-squark masses are shown in the green (dark gray) bands
of Fig.~\ref{fig:constraints}, with the specific generation of up squark
labeled inside the band.
The solutions with the additional ``*'' label, have the  quoted $\lambda_{i23}''$ coupling  absent
from the superpotential. However it is regenerated at order $\theta$
through a K\"ahler rotation~\cite{Choi:1996se} from the dominant
coupling still present in the superpotential.
As a result, again the negative solutions are excluded for the full range of squark
masses displayed in the figure. 
Moreover, the first two positive solutions are also excluded. 
In the figure, the gray region for $n_{\lambda''}\le-7$ is also shown.
In this case, the holomorphy of the superpotential forbids all the
$\lambda''$ terms and although they will be generated after $U(1)_H$
symmetry breaking via the K\"ahler potential~\cite{Giudice:1988yz},
these terms are suppressed by the additional factor
$m_{3/2}/M_P$~\cite{Sierra:2009zq}.
Therefore the LSP is very long-lived and the phenomenology at
colliders is expected to be the same as that in the MSSM.

Therefore, by demanding a $B$ violating model and imposing the
constraints on the $R$-parity breaking couplings, only positive
solutions for $n_{\lambda''}$ remain allowed giving rise to a clear
hierarchy between $\lambda''$ couplings, which have a direct impact on
the phenomenology of the LSP. 
The dominant coupling turns out to be $\lambda''_{323}$, a feature
shared with Refs.\cite{Csaki:2011ge,KerenZur:2012fr}.

\subsection{Dimension-five operators and proton decay}
So far the $U(1)_H$ symmetry has been used to forbid dimension-four lepton- number
violating couplings, in order to keep proton decay to a safe limit.
However,  proton decay mediated by $\lambda''$ couplings alone can occur in
scenarios with a gravitino lighter than a proton~\cite{Choi:1996nk},
leading to strong bounds on these couplings. Thus, by ensuring gravitino
masses greater than 1 GeV in these scenarios there will be no
contribution to the proton decay coming from a gravitino, 
which being the LSP can be also a dark matter canditate
\cite{Takayama:2000uz,Buchmuller:2007ui,Csaki:2011ge,Lola:2008bk,Lola:2007rw}.

On the other hand, there are also dimension-five lepton or$/$and baryon-
number violating couplings, which can induce proton decay.  Hence, it
is also necessary to check if these terms are also banned or
suppressed enough.
   
The nonrenormalizable dimension-five operators in the superpotential
$W_{5D}$ and K\"ahler potential $V_{5D}$ are given by
\cite{Allanach:2003eb, Barbier:2004ez, Ibanez:1991pr, Ibanez:1991hv}
\begin{align}\label{WD5}\nonumber
W_{D5}&=\frac{(\kappa_1)_{ijkl}}{M_P}\widehat{Q}_i\widehat{Q}_j\widehat{Q}_k\widehat{L}_l+\frac{(\kappa_2)_{ijkl}}{M_P}\widehat{u}_i\widehat{u}_j\widehat{d}_k\widehat{e}_l+\frac{(\kappa_3)_{ijk}}{M_P}\widehat{Q}_i\widehat{Q}_j\widehat{Q}_k\widehat{H}_d\\
&+\frac{(\kappa_4)_{ijk}}{M_P}\widehat{Q}_i\widehat{H}_d\widehat{u}_j\widehat{e}_k+\frac{(\kappa_5)_{ij}}{M_P}\widehat{L}_i\widehat{H}_u\widehat{L}_j\widehat{H}_u+\frac{(\kappa_6)_{i}}{M_P}\widehat{L}_i\widehat{H}_u\widehat{H}_d\widehat{H}_d,\\
V_{5D}&=\frac{(\kappa_7)_{ijk}}{M_P}\widehat{u}_i\widehat{d}^{*}_j\widehat{e}_k+\frac{(\kappa_8)_{i}}{M_P}\widehat{H}^{*}_u\widehat{H}_d\widehat{e}_i+\frac{(\kappa_9)_{ijk}}{M_P}\widehat{Q}_i\widehat{L}^{*}_j\widehat{u}_k+\frac{(\kappa_{10})_{ijk}}{M_P}\widehat{Q}_i\widehat{Q}_j\widehat{d}^{*}_k.
\end{align}

A review of the effect of these operators in the destabilization of the
proton is given in Ref.~\cite{Dreiner:2012ae}. 
In the present case of $B$ violation, we would guarantee a
sufficiently stable proton if the $B$ and $L$-violating  operators with couplings $\kappa_{1,2}$ 
and the $L$-violating operators with coupling $\kappa_{4,7,8,9}$ are forbidden\footnote{The constraints on the operator with coupling $\kappa_6$ are mild~\cite{Dreiner:2012ae}}.
The operator with coupling $\kappa_5$, $LH_uLH_u$, is not constrained
by proton decays because it violates the lepton number by two units.

The horizontal charges for all the dimension-5 operators are given
in Appendix~\ref{aped.B}.
Given the fractional values ​​needed for $n_i$ in order to get rid of
the dimension-four $L$ violating operators in Eq.~\eqref{po}, it turns
out that all dimension-five $L$-violating operators are also
automatically forbidden by the $U(1)_H$ symmetry see
Eqs.~(\ref{eq:D5BLNV1}), (~\ref{eq:D5BLNV2}) and (~\ref{eq:D5LNV}).
At this stage the $U(1)_H$ symmetry plays the same role as that of a lepton-
parity discrete symmetry \cite{Allanach:2003eb,Ibanez:1991pr,Ibanez:1991hv,Dreiner:2005rd}.

\section{Generation of neutrino masses}
\label{sec:RHN}
Although it is not required the $LH_uLH_u$ operator be forbidden
by $U(1)_H$ symmetry to ensure proton stability,
it is unavoidably prohibited because the
bilinear charges $n_i$ are not half-integers.  
Thus, the Majorana mass terms $\nu_L\nu_L$ are automatically forbidden.  
The same happens with lepton-parity symmetry, and also within the more general
approach of gauge discrete symmetries~\cite{Ibanez:1991pr,Ibanez:1991hv,Dreiner:2005rd},
for which the solutions than allow the $UDD$ operator
automatically forbid Majorana neutrinos.
The proposed solution in these kinds of frameworks is just to introduce
right-handed neutrinos $N$ with  their Majorana mass terms $NN$
forbidden, while keeping the Yukawa operators containing left- and
right-handed neutrinos still allowed, generating in this way Dirac neutrino
mass matrices~\cite{Luhn:2007gq}. 
When these ideas are applied to our case of horizontal symmetries, it is also necessary to explain the smallness of the neutrino
Yukawa couplings.
The introduction of three right-handed neutrinos $N_{i}$ $(i=1,2,3)$ allows us to give 
Dirac masses to neutrinos by assigning fractional and not half-integer
$H$-charges to $N_{i}$, such that the $NN$ terms remain forbidden.

Let us paramatrize the bilinear $H$ charges as $n_2=n_1+\alpha$,
$n_3=n_1+\beta$ and for right-handed neutrinos: $N_2=N_1+\epsilon$
and $N_3=N_1+\rho$. The neutrino Dirac mass matrix reads
\begin{align}
 M_{\nu}\sim 
v_u\theta^{\beta+\rho+n_1+N_1}
\left(
\begin{array}{ccc}
 \theta^{-\beta-\rho } & \theta^{\epsilon -\beta-\rho} & \theta^{-\beta} \\
 \theta^{\alpha -\beta-\rho } & \theta^{\alpha +\epsilon-\beta -\rho} & \theta^{\alpha-\beta} \\
 \theta^{-\rho} & \theta^{\epsilon-\rho } & 1
\end{array}
\right),
\end{align}
where  $v_{u}$ is the vacuum expectation value developed by the up-type Higgs field. 
From Eq.~(\ref{nlpp}) we obtain $n_1=\frac{1}{3}\left(1-\alpha -\beta
  +3 n_{\lambda ''}-3 x\right)$. Motivated by the recent results of a
large value for $\theta_{13}$
\cite{Abe:2011sj,Abe:2011fz,An:2012eh,Ahn:2012nd}, which support those
models based on a anarchical neutrino mass matrix
\cite{Hall:1999sn,Haba:2000be,deGouvea:2003xe,deGouvea:2012ac}, it is
convenient to choose $\alpha=\beta=\epsilon=\rho=0$ and
$\beta+\rho+n_1+N_1=n_{Y_\nu}$ with $n_{Y_\nu}$ being an integer and $n_{Y_\nu}\ge 16$ 
in order to generate a neutrino Yukawa coupling 
$Y_\nu\lesssim10^{-11}$. 
It is worth stressing that since $n_1$ cannot be an integer, the
$\mu\tau$ anarchical texture with $\alpha=\beta=-1$
\cite{Dreiner:2003yr, Altarelli:2002sg, Buchmuller:2011tm, Altarelli:2012ia}
is not allowed.
However, other textures can be accommodated in our model 
\cite{Altarelli:2012ia}, such as pseud- $\mu\tau$ anarchy
($\alpha=\beta=\epsilon=\rho=-2$) and the hierarchical texture
($\alpha=\epsilon=-1,\, \beta=\rho=-2$).
An immediate consequence of the anarchy assumption is that the
bilinear charges are equal and are set to $n_i=n_{\lambda
  ''}-x+\frac{1}{3}$, being clearly noninteger numbers.
The $H$-charges that allow us to obtain a self-consistent framework with
the requirements mentioned above are shown in Table~\ref{tab:Hc3}. 
It is remarkable that when explaining the neutrino Yukawa couplings
$Y_\nu$, a lower bound on $n_{\lambda''}\ge 6$ emerges, which leads
to deep implications on the phenomenology of the model (see the next
Section).

\begin{table}[t]
  \centering
  \begin{tabular}{ccccccccc}\hline
    $x$            & 0 & 0 & 0 & 0 & 1 & 0 & 1 & 2 \\\hline 
    $n_{\lambda''}$ & 6 & 7 & 8 & 9 & 9 & 10 & 10& 10\\
    $n_i$ & 19/3 & 22/3 & $25/3$ & $28/3$ & $25/3$ & $31/3$ & $28/3$&$25/3$  \\
    $N_i$ & $29/3$ & $26/3$ & $29/3$ & $26/3$ & $29/3$ &$23/3$ & $26/3$&29/3   \\\hline 
      \end{tabular}
      \caption{Some sets of $H$-charge allowing a  self-consistent framework of 
        $R$-parity breaking with $B$ violation and Dirac neutrinos.}
  \label{tab:Hc3}
\end{table}
\subsection{Majorana neutrinos}
It is worth mentioning that it is also possible to have Majorana
neutrinos if in addition to the right-handed neutrinos we include in the
model a second flavon\footnote{For a model with several flavons see
  Ref.\cite{Jack:2003pb}}, $\psi$, with fractional\footnote{A scenario
  with Majorana neutrinos and nonanomalous $U(1)_H$ symmetry, which
  is spontaneously broken by two flavons with opposite $H$-charge +1
  and -1 was obtained in Ref.\cite{Eyal:1999gq}.} $H$-charge and with a
vacuum expectation value approximately equal to $\theta$.
The horizontal charges of these superfields are fixed by new invariant
diagrams coming from Dirac and Majorana mass terms.
 
In this way, the $H$-charge of $\psi$ must be such that it does not get
coupled to $L$-violating operators. Therefore, the respective total
$H$-charge of the full $L$ violating operator would be either
fractional and therefore forbidden, or negative and sufficiently
suppressed.

The introduction of an additional flavon field could spoil the proton
stability since $H$-invariant terms can be obtained by coupling a
large number of $\psi$ flavons to dangerous operators.
Therefore  it is  mandatory  to ensure  that  $L$ violating  bilinear,
dimension-four  and  dimension-five operators  are  generated  through the GM
mechanism or have a large Froggatt-Nielsen suppression. 
The $H$-charges that allow us to obtain Majorana neutrinos with
the requirements mentioned above, are shown in Table~\ref{tab:Hc}.  
To illustrate this point, let us consider the first solution given in Table~\ref{tab:Hc}. 
For that set of $H$-charges, we have found that the minimum suppression that is achieved 
for dimension-four and -five operators is 
$\widehat{L}_1\widehat{Q}_1\widehat{D}_1: m_{3/2}\theta^{21}/M_P$ and 
$\widehat{u}_1\widehat{u}_2\widehat{d}_1\widehat{e}_1: m_{3/2}\theta^2/M_P^2$, 
which is enough to satisfy the constraints coming from proton decay.

\begin{table}[t]
  \centering
  \begin{tabular}{ccccccccc}\hline
    $x$            & 1 & 1 & 1 & 2 & 2 & 2 & 3 & 3 \\\hline 
    $n_{\lambda''}$ & 5 & 6 & 7 & 6 & 7 & 8 & 8 & 9 \\
    $n_i$ & $13/3$ & $16/3$ & $19/3$ & $13/3$ & $16/3$ & $19/3$ & $16/3$ & $19/3$ \\
    $\psi$ & $-47/6$ & $-53/6$ & $-59/6$ &$-47/6$ & $-53/6$ & $-59/6$ &$-53/6$ & $-59/6$  \\\hline 
      \end{tabular}
      \caption{Sets of $H$-charges that allow having Majorana neutrinos with $H(N_i)=7/2$. 
        For this scenario there is no lower bound on $n_{\lambda''}$.}
  \label{tab:Hc}
\end{table}

Henceforth, we will combine the solutions allowed by the
experimental constraints on $R$-parity breaking couplings discussed in
Sec.~\ref{sec:BNVmodel}, with the restrictions to obtain Dirac
neutrinos, and therefore we will only consider solutions with
$n_{\lambda''}\ge 6$.

\section{Implications on collider searches}
\label{sec:implications}
From a collider physics point of view, there are two main differences
between the models with and without $R$-parity conservation.
When $R$-parity conservation is assumed, the production
of supersymmetric particles is in pairs, and the LSP is stable leading
to missing energy signatures in the detectors.
On the other hand, $R$-parity violation allows for the single
production of supersymmetric particles and the decay of the LSP
involving jets or/and leptons.
The $R$-parity breaking and $B$ violating operators induce LSP decay
directly or indirectly to quarks, including the top if LSP is
sufficiently massive\footnote{If a supersymmetric partner of some SM
  particle is the nest-to-the-lightest-supersymmtric particle with the gravitino as the LSP, our
  phenomenological results would not change.}.
Given that the LSP is no longer stable due to R-parity violation, in
principle the LSP can be any supersymmetric particle
\cite{Dreiner:1997uz, Barbier:2004ez,Dreiner:2008ca}.
For recent phenomenological studies in supersymmetric scenarios with
$R$-breaking through $B$ violating, see, {\it e.g.} Refs.
\cite{Butterworth:2009qa,Allanach:2012vj,Brust:2012uf,Asano:2012gj,Curtin:2012rm,Lola:2008bk,Dreiner:2008ca,Evans:2012bf,Berger:2012mm,Carpenter:2008sy,Dreiner:2012wm,
Franceschini:2012za,Bomark:2011ye,Choudhury:2011ve,Desai:2010sq,Kilic:2011sr,Carpenter:2007zz,Kaplan:2007ap,Berger:2013sir}
and in particular, Refs. \cite{Csaki:2011ge,KerenZur:2012fr}.

The phenomenology of the model at LHC is basically the same studied in
the SSM with minimal flavor violation (MFV)~\cite{Csaki:2011ge} and partial compositeness \cite{KerenZur:2012fr}. 
In fact, in Ref.~\cite{Csaki:2011ge} they also get a hierarchy in which the third-generation couplings
dominate with fixed ratios between them.
Fixing the expansion parameter as $\theta=0.22$, their set of
$R$-parity breaking parameters can be written as
\begin{align}
\label{eq:Grosslpp}
 \begin{pmatrix}
    \lambda_{112}''&\lambda_{212}''&\lambda_{312}''\\
    \lambda_{113}''&\lambda_{213}''&\lambda_{313}''\\
    \lambda_{123}''&\lambda_{223}''&\lambda_{323}''\\
  \end{pmatrix}\approx&
 \tan^2\beta_{\text{MFV}}\begin{pmatrix}
   \theta^{24} & \theta^{18} & \theta^{13}\\
   \theta^{19} & \theta^{14} & \theta^{12}\\
   \theta^{16} & \theta^{13} & \theta^{11}\\
  \end{pmatrix}=
 \theta^{n_{\text{MFV}}}\begin{pmatrix}
   \theta^{13} & \theta^{7} & \theta^2\\
   \theta^8 & \theta^3 & \theta\\
   \theta^5 & \theta^2 & 1\\
  \end{pmatrix}\,,
\end{align}
with $\theta^{n_{\text{MFV}}}=\theta^{11}\tan^2\beta_{\text{MFV}}$. 
Comparing with Eq.~\eqref{eq:Hierlpp}, we can see that the set of
predicted couplings until order $\theta^{n_{\text{MFV}}+3}$ is
basically the same as in our case (with the exception of their
$\lambda_{312}''$ which has an additional suppression factor of
$\theta$).
Therefore, the phenomenology of both theories for $R$-parity violation
should be the same at the LHC. 
In fact, the phenomenology of Ref.~\cite{Csaki:2011ge} for the leading
couplings was analyzed in detail at the  LHC with the results
presented as function of $\tan\beta_{\text{MFV}}$.
The specific values at
$\tan\beta_{\text{MFV}}\approx(44.5,20.7,9.7,4.6,2.1)$ in several
plots of Ref.~\cite{Csaki:2011ge} correspond to the discrete set of solutions
$n_{\lambda''}=(6,7,8,9,10)$ respectively, in our model. 
In particular several plots there, they explore the decay length ($c\tau$)
for LSP masses in the range of $100-800\ \text{GeV}$.
When the stop is the LSP for example, displaced vertices (DV) are
expected for $n_{\lambda''}=10$.
For a sbottom LSP it is possible to have DV for $n_{\lambda''}=9,10$,
while the three-body decays of a LSP neutralino could generate DV for
$n_{\lambda''}=8,9,10$.
In the same vein, because decays of the stau LSP involves four
particles in the final state, DV are expected for $n_{\lambda''}\ge
6$.

Recent phenomenological analysis in $R$-parity breaking
trough $UDD$ operators has focused on prompt decay for stops and
sbottoms~\cite{Csaki:2011ge,Franceschini:2012za,Bhattacherjee:2013gr,Csaki:2013we}.
However, the experimental results about DV at the LHC
are, in general not directly applicable to these kinds of models, because
high $p_T$ leptons are required for trigger the
events~\cite{Aad:2011zb,Aad:2012zx,Chatrchyan:2012jna}, and to be part of the
DV~\cite{Aad:2012zx,Chatrchyan:2012jna}. 
We assume in the discussion below that pure hadronic DV are still
compatible with light squarks and gluinos.

Regarding collider searches, a pair produced gluino with a prompt decay to three
jets has been searched by CDF~\cite{Aaltonen:2011sg}, CMS
\cite{Chatrchyan:2011cj,Chatrchyan:2012uxa} and ATLAS~\cite{ATLAS:2012dp}.%
\footnote{In this analysis all the superpartners except for the
  gluinos are decoupled, and some reinterpretation would be needed to
  apply the results to a more generic SUSY spectrum.}
CMS results constrain the gluino mass to be in the ranges
$144<m_{\tilde{g}}<200$ GeV or $m_{\tilde{g}}>460$ GeV.
However, ATLAS already excludes gluino masses up to
$m_{\tilde{g}}\lesssim666$ GeV.
In general, these bounds do not apply when the gluino is not the
LSP~\cite{Csaki:2011ge,Bhattacherjee:2013gr}.  
On the other hand, CDF~\cite{Aaltonen:2008dn}, ATLAS
~\cite{Aad:2011yh,ATLAS:2012ds} and CMS~\cite{Chatrchyan:2013izb} also
have performed searches for pair production of dijet resonances in
four-jet events without putting appreciable constraints on stops
decaying to dijets.
Therefore, the already analyzed data at the LHC still allow for low
squarks and gluinos in scenarios with $R$-parity breaking through
$B$-violating couplings~\cite{Evans:2012bf,Bhattacherjee:2013gr} 

We have seen that both this single-flavon horizontal (SFH) and the MFV models, lead to a
realistic and predictive framework which could be more easily probed
at LHC than some {\it ad hoc} version or $R$-parity breaking with $B$
violation.  
In fact, recently in~\cite{Berger:2013sir} the CMS results on searches for
new physics in events with same-sign dileptons and $b$ jets
~\cite{Chatrchyan:2012paa} have been recasting in a simplified version of the
$R$-parity breaking MFV model where it is assumed one spectrum with
only two light states: a gluino and a stop. 
All other SUSY particles are assumed to be either too heavy or too
weakly coupled to be relevant at the LHC. 
Furthermore, the stop is assumed to be the LSP, and
$m_{\tilde{g}}>m_{\tilde{t}}+m_t$.\footnote{As a consequence the gluino branching to stop-top is equal to 1.}  
Under these conditions they are able to set a lower bound on the gluino
mass about $800\ \text{GeV}$ at 95\% of confidence level.\footnote{The
  obtained lower bound only apply if the gluino is a Majorana
  particle.}
The same bound could apply to the SFH model with
$R$-parity breaking presented in this work.

In order to really probe this single-flavon horizontal (or the MFV) 
$R$-parity breaking model, the full textures in Eq.~\eqref{eq:Hlpp} or ~\eqref{eq:Grosslpp}
should be probed.
However, relations between different branching ratios could be
measured only in $e^+e^-$ colliders.
In a stop LSP scenario, it can decay directly into two down quarks of
different generations through the $\lambda''_{3jk}$ coupling.  In this
case, the hierarchy between $\lambda''$ couplings allows for estimate
several fractions of branchings, e.g.  $\operatorname{Br}(\tilde t\to
\bar{s}\bar{b})/\operatorname{Br}(\tilde t\to
\bar{d}\bar{s})/\sim\theta^2$.
A sbottom LSP, with a mass larger than the top mass, may show the
clear hierarchy $\operatorname{Br}(\tilde b\to
\bar{t}\bar{s})/\operatorname{Br}(\tilde b\to
\bar{c}\bar{s})/\sim\theta^4$.
For a neutralino LSP with $m_{\tilde\chi}^0>m_t$, the dominant
coupling $\lambda''_{323}$ entails $\operatorname{Br}(\tilde\chi\to t
d b)/\operatorname{Br}(\tilde\chi\to t s
b)\sim\operatorname{Br}(\tilde\chi\to t d
s)/\operatorname{Br}(\tilde\chi\to t s b)\sim\theta^2$ and
$\operatorname{Br}(\tilde\chi\to c s
b)/\operatorname{Br}(\tilde\chi\to t s b)\sim\theta^4$.
For the case $m_{\tilde\chi}<m_t$ the main neutralino decay is then
controlled by $\lambda''_{223}$, and  will produce charm quarks with
ratios of branching ratios given by $\operatorname{Br}(\tilde\chi\to c
d b)/\operatorname{Br}(\tilde\chi\to c s
b)\sim\operatorname{Br}(\tilde\chi\to c d
s)/\operatorname{Br}(\tilde\chi\to c s b)\sim\theta^2$.

\chapter{Neutrino masses in $SU(5)\times U(1)_H$ with adjoint flavons}\label{cap3}

We present a $SU(5)\times U(1)_H$ supersymmetric model for neutrino masses and mixings that implements the seesaw mechanism by means of the heavy $SU(2)$ singlets and triplets states contained in three adjoint of $SU(5)$.  We discuss how Abelian $U(1)_H$ symmetries can naturally yield non-hierarchical light neutrinos even when the heavy states are strongly hierarchical, and how it can also ensure that $R$--parity arises as an exact accidental symmetry.  By assigning two flavons that break $U(1)_H$ to the adjoint representation of $SU(5)$ and assuming universality for all the fundamental couplings,the coefficients of the effective Yukawa and Majorana mass operators become calculable in terms of group theoretical quantities.  There is a single free parameter in the model, however, at leading order the structure of the light neutrinos mass matrix is determined in a parameter independent way.

\section{Theoretical framework}
\subsection{Same sign and both signs Abelian charges}
\label{sec:signs}

Sometimes symmetry considerations are sufficient to determine
univocally the structure of the low energy operators, however, other
times a detailed knowledge of the full high energy theory is needed.
Let us consider for example a $U(1)_H$ symmetry and assume that all
the heavy and light states have charges of the same sign, say
positive. Then a single spurion $\epsilon_{-1}$ with a negative unit
charge is involved in the construction of all $U(1)_H$ (formally)
invariant operators. Let us consider the ${\rm dim}=5$ seesaw operator
$ \mathcal{ L}_{D5} \sim - \frac{g_{\alpha\beta} }{2M}\left(\bar\ell_\alpha H\right) \left( H^T\ell^c_\beta\right) $, where
$\ell_\alpha$ are the lepton doublets and $H$ is the Higgs field, that
for simplicity we take neutral under the Abelian symmetry $F(H)=0$.
Since the only spurion useful to construct (formally) invariant
operators is $\epsilon_{-1}$, one can easily convince himself that the
structure of $g_{\alpha\beta}$, and thus the structure of the light neutrino
mass matrix, is univocally determined by the $F$ charges of the
light leptons as: $g_{\alpha\beta} \sim \epsilon_{-1}^{
  F(\ell_\alpha)+ F(\ell_\beta)}$, while the $F$-charges of whatever
heavy states of mass $\sim M$ are inducing the effective operator are
irrelevant.\footnote{It should be remarked that, contrary to what is
  sometimes stated, Abelian $U(1)_H$ symmetries allow to arrange very
  easily for non-hierarchical light neutrinos together with strongly
  hierarchical heavy neutrinos (as are often preferred in
  leptogenesis) by simply choosing $F(\ell_\alpha)=F(\ell)$ for all
  $\alpha$, and $ F(N_1)\gg F(N_2)\gg F(N_3)$.}  We can conclude that
in this case one does not need to consider the details of the high
energy theory, since the structure of the low energy effective
operators can be straightforwardly read off from the charges of the
light states.

However, if we allow for $U(1)_H$ charges of both signs, then both
symmetry breaking spurions $\epsilon_{-1}= \epsilon_{+1}= \epsilon$
are relevant. This implies that naive charge counting applied to the
low energy effective operators is unreliable, since basically a factor
$\epsilon^n$, as estimated in the low energy theory, could correspond
instead to $\epsilon^{n+m}_{+1} \cdot \epsilon^m_{-1} \sim
\epsilon^{n+2m}$.  Clearly the naive estimate can result in a
completely different (and wrong) structure with respect to the one
effectively generated by the high energy theory.  We illustrate this
with a simple example: let us take two lepton doublets with charges
$F(\ell_1)= -F(\ell_2)= +1$ and again $F(H)=0$.  The structure of the
light neutrino mass matrix read off from the lepton doublets charges
would be given by the low energy coefficient:
\begin{equation}                     
  \label{eq:mnu1}                                                 
   g_{\alpha\beta}  
\sim \begin{pmatrix}  
\epsilon^2 & 1 \\ 1 & \epsilon^2 \end{pmatrix} \,.
\end{equation}
This corresponds to a pair of quasi degenerate (pseudo-Dirac) light
neutrinos.

Now, let us assume that the fundamental high energy (seesaw) theory
has two right handed  neutrinos with charges $F(N_{1,2})=+1$.
For the heavy mass matrix $M_N$, its inverse, and for the Yukawa coupling
$Y_{\alpha i}\bar\ell_\alpha N_i$ we obtain:
\begin{align}
  \label{eq:MN}
M_N &\sim \epsilon^2               
\begin{pmatrix} 
1&1\\1&1
\end{pmatrix}\,,&
M_N^{-1} &\sim   \epsilon^{-2}
\begin{pmatrix}1&1\\ 1&1\end{pmatrix}\,,&
Y &\sim  
\begin{pmatrix}1&1\\ \epsilon^2&\epsilon^2\end{pmatrix}\,.
\end{align}
The resulting effective low energy coefficient  is: 
\begin{equation}                                                  
 \label{eq:mnu2}                                                 
g_{\alpha\beta}  \sim                                                    
Y M_N^{-1}Y^T  \sim  
 \epsilon^{-2}
\begin{pmatrix}
1 & \epsilon^2 \\
\epsilon^2 & \epsilon^4
\end{pmatrix}
\,,
\end{equation}
which (for $\epsilon\ll 1$) corresponds to very hierarchical and
mildly mixed light neutrinos, that is a completely different result
from the previous one.   

The model we are going to describe in this paper requires fermions
with charges of both signs, as well as a pair of positively and
negatively charged spurions. Therefore a detailed knowledge of the
high energy theory is mandatory, and accordingly we will explicitly
describe all its relevant aspects.

\subsection{Outline of the $SU(5)\times U(1)_H$ model}

We assume that at the fundamental level all the Yukawa couplings are
universal, and that all the heavy messengers states carrying $U(1)_H$
charges have the same mass, as it would happen if the masses are
generated by the vacuum expectation values (vev) of some singlet
scalar.  With these assumptions, the only free parameter of the model
is the ratio between the vacuum expectation value of the flavons and
the mass of the heavy vectorlike FN fields. This parameter is
responsible for the fermion mass hierarchy, and all the remaining
features of the mass spectrum are calculable in terms of group
theoretical coefficients.  More precisely, in our model the flavor
symmetry is broken by vevs of scalar fields $\langle
\Sigma_{\pm}\rangle$ in the $\mathbf{24}$--dimensional adjoint
representation of $SU(5)$, where the subscripts refer to the values
$\pm 1$ of the $U(1)_H$ charges that set the normalization for all the
other charges. The vevs
$\langle\Sigma_+\rangle=\langle\Sigma_-\rangle=V_a$ with
$V_a=V\cdot\operatorname{diag}(2,2,2,-3,-3)/\sqrt{60}$ are also
responsible for breaking the GUT symmetry down to the
electroweak--color gauge group. The size of the order parameters
breaking the flavor symmetry is then $\epsilon=V/M$ where $M$ is the
common mass of the heavy FN vectorlike fields. This symmetry breaking
scheme has two important consequences: power suppression in
$\epsilon$ appear with coefficients related to the different entries
in $V_a$, and the FN fields are not restricted to the $\mathbf{5}$,
$\overline{\mathbf{5}}$, or $\mathbf{10}$, $\overline{\mathbf{10}}$,
multiplets as is the case when the $U(1)_H$ breaking is triggered by
singlet flavons~\cite{Aristizabal:2003zn,Duque:2008ah}. 

The model studied in~\cite{Duque:2008ah} adopted this same scheme, and
yields a viable phenomenology, since it produces quark masses and
mixings and charged lepton masses that are in agreement with the data.
The $U(1)_H$ charge assignments of the model yield $U(1)_H$ mixed
anomalies, that are canceled trough the Green-Schwartz
mechanism~\cite{Green:1984sg}. The values of the charges are
determined only modulo an overall rescaling, that may be appropriately
chosen in order to forbid baryon and lepton number violating
couplings. However, with the choice of charges adopted
in~\cite{Duque:2008ah}, both $\Delta L=1$ and $\Delta L=2$ violating
operators were forbidden, and thus the seesaw mechanism could not be
embedded in the model. In order to avoid this unpleasant feature, in
this work we explore the possibility of forbidding just the $\Delta
L=1$ operators while allowing the $\Delta L=2$ seesaw operator for
neutrino masses.  We will show that by means of a suitable choice of
the $F$ charges, the seesaw mechanism can be implemented, and one can
obtain neutrino masses and mixings in agreement with oscillation data,
while $\Delta L=1$ and $\Delta B\neq 0$ (and thus $R$--parity
violating) operators are forbidden at all orders by virtue of the
$F$-charges.  Moreover, the scale of the heavy seesaw neutral fermions
remains fixed, and lies a few order of magnitude below the GUT scale,
and is of the right order to allow the generation of the baryon
asymmetry through leptogenesis.

\subsection{Charge assignments}

The  $F$ charges have to satisfy some specific
requirements in order to yield a viable phenomenology.  In the
following we denote for simplicity the various $F$ charges with the
same label denoting the corresponding $SU(5)$ multiplet.  To
allow a Higgsino $\mu$--term at tree level, we must require
\begin{align}
\label{eq:mu}
  \overline{\mathbf{5}}^{\phi_d}+\mathbf{5}_{\phi_u}=0\,,
\end{align}
where $ \overline{\mathbf{5}}^{\phi_d},\,\mathbf{5}_{\phi_u}$ denote
the $F$-charges of the 
chiral multiplets containing the $SU(2)$  Higgs doublets 
$\phi_d,\,\phi_u$. 
It is easy to see that with the constraint~(\ref{eq:mu}) the
overall charge of the Yukawa operators for the charged fermion masses
$ \mathbf{10}_{I} \mathbf{\bar{5}}_{J} \mathbf{\bar{5}}^{{\phi}_{d}}$
and $ \mathbf{10}_{I} \mathbf{10}_{J} \mathbf{{5}}_{{\phi}_{u}}$, that
are even under $R$--parity, are invariant under the charge
redefinitions~\cite{Duque:2008ah}:
\begin{align}
\label{eq:shift}
      \mathbf{\bar{5}}_{I}&\to \mathbf{\bar{5}}_{I} + a_n\\ \nonumber
    \mathbf{10}_{I} &\to  \mathbf{10}_{I} -\frac{a_n}{3}\\ \nonumber
    \mathbf{\bar{5}}^{{\phi}_{d}} &\to
    \mathbf{\bar{5}}^{{\phi}_{d}} - \frac{2a_n}{3}\\ \nonumber
    \mathbf{5}_{{\phi}_{u}}&\to \mathbf{5}_{{\phi}_{u}} +\frac{2a_n}{3}\,,
\end{align}
where $I=1,2,3$ is a generation index, and $a_n$ is an arbitrary
parameter that can be used to redefine the charges.  Assuming
$\mathbf{5}_{\phi_u}= 0$,
then the anomalous solution that
was chosen in ref.~\cite{Duque:2008ah} can be written as
\begin{align}
  \label{eq:1}
\mathbf{5}_{\phi_u}= \overline{\mathbf{5}}^{\phi_d} = &0\,, &&
\overline{\mathbf{5}}_I=2^I-7\,, &&\mathbf{10}_I=3-I\,.
\end{align}

Starting from a set of integer charges, and redefining 
this set by means of the shift 
Eq.~(\ref{eq:shift}) with 
\begin{align}
  \label{eq:2}
  a_n=-\frac{3}{2}\left( \frac{2n}{5} +1\right)\,,
\end{align}
where $n$ is an integer, it is easy to see that the $R$--parity
violating operators $\mathbf{10}_{I} \mathbf{\bar{5}}_{J}
\mathbf{\bar{5}}_{K}$ and $\mathbf{\bar{5}}_{I}
\mathbf{{5}}_{{\phi}_{u}}$ have half--odd--integer charges, and hence
are forbidden at all orders by the $U(1)_H$ symmetry. 

To generate neutrino masses, we now introduce three heavy multiplets
$\mathbf{N}_I$ $(I=1,2,3$) with half--odd--integer $F$--charges, that
we assume corresponding to adjoint representations $\mathbf{24}$.  The
adjoint of $SU(5)$ contains two types of $SU(2)$ multiplets that can
induce at low energy the dimension five Weinberg
operator~\cite{Weinberg:1979sa}: one $SU(2)\times U(1)\times SU(3)$
singlet that allows to implement the usual type I seesaw, and one
$U(1)\times SU(3)$ singlet but $SU(2)$ triplet giving rise to a type
III seesaw~\cite{Bajc:2006ia,Bajc:2007zf,Biggio:2010me}. Contributions
from these two types of multiplets unavoidably come together, so that
by assigning `right handed neutrinos' to the $\mathbf{24}$ of $SU(5)$
one necessarily ends up with a type I+III seesaw.\footnote{We thank
  the referee for bringing this point to our attention.}
%
%
This slightly more complicated seesaw structure is not crucial for our
construction, but we still keep track of it for a matter of
consistency.

The half--odd--integer charges of the new states, after the charges of
the other fields have been shifted according to Eqs.~(\ref{eq:shift})
and (\ref{eq:2}), can be parameterized as
\begin{align}
\label{eq:3}
N_I=\frac{2m_I + 1}{2}\,,
\end{align}
where $m_I$ are integers. The  {\it effective} superpotential terms
that give rise to the seesaw are
\begin{align}
\label{eq:4}
  W_{\text{seesaw}}=Y_\nu^{I J}\,\overline{\mathbf{5}}_{I}\,\mathbf{5}_{\phi_u}\,\mathbf{N}_{J}
+\tfrac{1}{2}M_R^{I J} \mathbf{N}_I \mathbf{N}_J\,.
\end{align}
The coefficient $Y_\nu^{I J}$ of the Dirac operator in
Eq.~\eqref{eq:4} is determined by the following sums of $F$--charges:
\begin{align}
\overline{\mathbf{5}}_{I}+\mathbf{5}_{\phi_u}+\mathbf{N}_{J}
=&2^I-7+a_n+2a_n/3+N_J\nonumber\\
=&2^I-9+n+m_J\,.\nonumber
\end{align}
Explicitly:
\begin{align}
\label{eq:5}
  F(\overline{\mathbf{5}}_{I}\,\mathbf{5}_{\phi_u}\,\mathbf{N}_{J})=&
\begin{pmatrix}
    -7-n+m_1 &  -7-n+m_2 & -7-n+m_3 \\
   -5-n+m_1 &  -5-n+m_2 & -5-n+m_3 \\
   -1-n+m_1 &  -1-n+m_2 & -1-n+m_3 \\
\end{pmatrix}.
\end{align}
For the mass operator of  the adjoint neutrinos 
we have the following (integer) $F$--charges
\begin{align}
\label{eq:6}
   \mathbf{N}_{I}+\mathbf{N}_{J}=&1+m_I+m_J\,,\nonumber\\
  F(\mathbf{N}_I\,\mathbf{N}_J) =&\begin{pmatrix}
    1+ 2m_1   &1+ m_1+m_2 &1+m_1+m_3  \\
    1+ m_1+m_2 &1+ 2m_2  &1+m_2+m_3\\
    1+ m_1+m_3 &1+ m_2+m_3  &1+2m_3\\
   \end{pmatrix}.
\end{align}
The light neutrino
mass matrix is then obtained from the seesaw formula
\begin{align}
  \label{eq:7}
  M_\nu \approx& - {v^2}\sin^2\beta\,Y_\nu\,M_R^{-1}\,Y^T_{\nu}\,,
\end{align}
where $v=175\,$GeV, and it is left understood that in Eq.~\eqref{eq:7}
the contributions of the $SU(2)$ singlets and triplets are both summed
up.
As is implied by the FN mechanism, the order of magnitude of the
entries in $Y_\nu$ and $M_R$ is determined by the corresponding values
of the sums of $F$ charges Eqs.~(\ref{eq:5}) and
(\ref{eq:6}) as:
\begin{align}
  \label{eq:8}
  Y_\nu^{I J}&\sim\epsilon^{|\overline{\mathbf{5}}_{I}+\mathbf{5}_{\phi_u}+\mathbf{N}_{J} |}
\nonumber\\
M_R^{I J}&\sim M 
\cdot\epsilon^{|\mathbf{N}_{I}+\mathbf{N}_{J}|} = 
V \cdot\epsilon^{|\mathbf{N}_{I}+\mathbf{N}_{J}|-1}\,.
 \end{align}
 where in the second relation $M$ is the mass of the FN messengers
 fields and in the last equality we have used $M=\epsilon^{-1} V$.
 Note that since we have two flavon multiplets $\Sigma_{\pm}$ with
 opposite charges, the horizontal symmetry allows for operators with
 charges of both signs, and hence the exponents of the symmetry
 breaking parameter $\epsilon$ in Eq.~\eqref{eq:8} must be given in
 terms of the absolute values of the sum of charges.  In FN models
 only the order of magnitude of the entries in Eq.~\eqref{eq:8} are
 determined, and it is generally assumed that non-hierarchical order
 one coefficients multiply each entry.  However, in our model the
 assumption of universality for the fundamental Yukawa couplings has
 been made in order to avoid arbitrary $\mathcal{O}(1)$ numbers of
 unspecified origin.\footnote{This condition excludes the simple (and
   often used) charge assignments in which there are two zero
   eigenvalues in the light neutrino mass matrix, as
   in~\cite{Dreiner:2003yr,Chen:2008tc}.}  The coefficients
 multiplying each entry in Eq.(\ref{eq:8}) can be in fact computed
 with the same technique introduced in~\cite{Duque:2008ah} for
 computing the down-quark and charged lepton masses.  In summary, the
 order of magnitude of the various entries in $M_\nu$ is determined by
 the appropriate powers of the small factor $\epsilon$ while, as we
 will see, the details of the mass spectrum are determined by
 non-hierarchical computable group theoretical coefficients, that only
 depend on the way the heavy FN states are assigned to $SU(5)$
 representations.

\subsection{Coefficients of the Dirac and Majorana effective
   operators}

In this section we analyze the contributions of different effective
operators to $Y_\nu$ and to $M_R$, showing that a phenomenologically
acceptable structure, able to reproduce (approximately) the correct
mass ratios and to give reasonable neutrino mixing angles can be
obtained.

\begin{figure}[t!]
   \centering

\hspace{0.5cm}\includegraphics[scale=0.75]{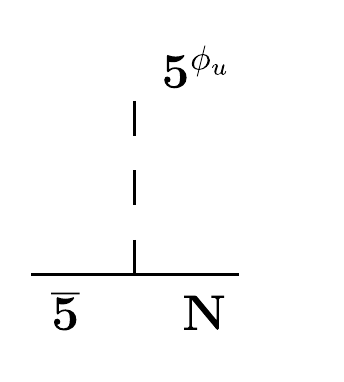} 
\hspace{2.3cm}\includegraphics[scale=0.75]{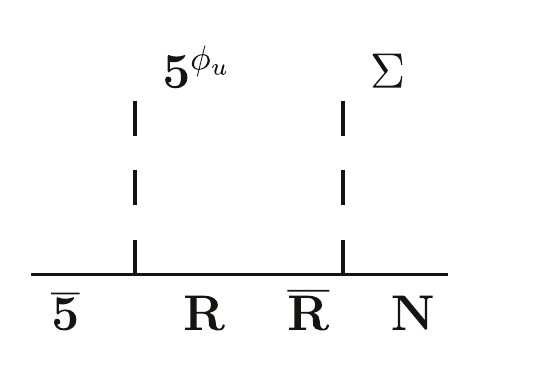}   
\hspace{1.7cm}\includegraphics[scale=0.75]{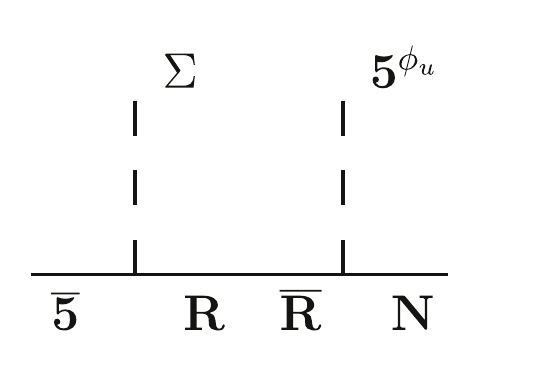} \\ [-5pt]
\leftline{\hspace{2.4cm}(a)\hspace{5.5cm}(b)\hspace{5.5cm}(c)} 

\vspace{10pt}

\includegraphics[scale=0.75]{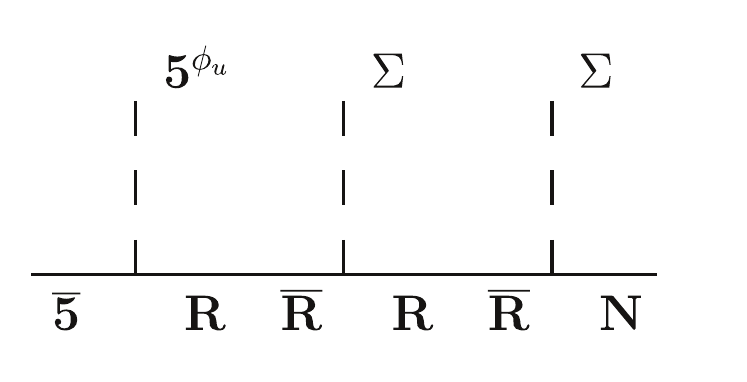}   
\includegraphics[scale=0.75]{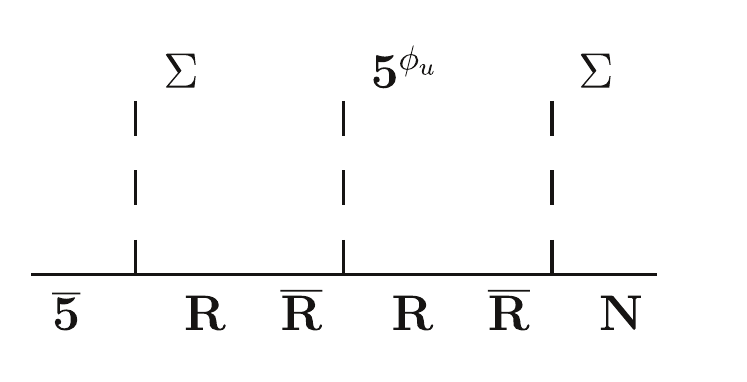} 
\includegraphics[scale=0.75]{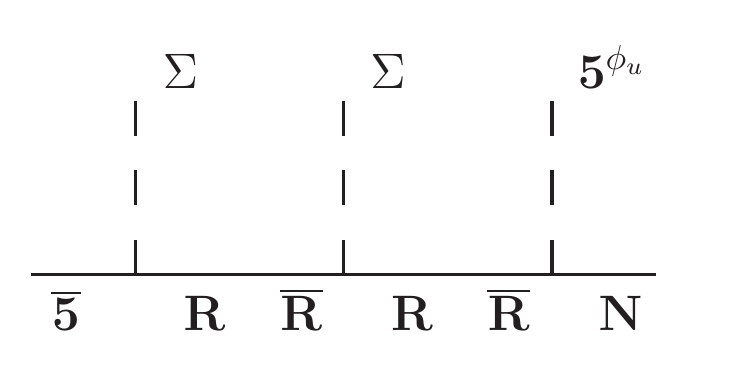} \\ [-5pt]
\leftline{\hspace{2.5cm} (d) \hspace{5.1cm} (e) \hspace{5.0cm} (f)}

\caption{ Diagrams contributing to $Y_\nu$ at different orders. The
  lowest order coefficient corresponding to diagram (a) is $y_0=3$.
  Diagrams (b)--(c) contribute at $\mathcal{ O}(\epsilon^1)$ and yield the
  coefficients $y_1$ in the second column in
  Table~\ref{tab:1}. Diagrams (d)--(f) contribute at $\mathcal
    {O}(\epsilon^2)$ and give the coefficients $y_2$ in the fourth
  column of the table.}
  \label{fig:1}
\end{figure}

We assume that a large number of vectorlike FN fields exist in various
$SU(5)$ representations.  Since we assign the heavy Majorana neutrinos
to the adjoint $\mathbf{N}$, the possible FN field
representations $\mathbf{R}$ can be identified starting from the
following tensor products involving the representations of the
fields in the external lines (see the diagrams in Fig.~\ref{fig:1}):
\begin{align}
  \label{eq:9}
\mathbf{ \overline{5} \otimes {5_{\phi_u}} } =& \mathbf{
{1}\oplus{24} } \,, \\
\mathbf{\overline{5}} \>\> \mathbf{\otimes} \>\> \mathbf{\Sigma}\  =& \mathbf{
\overline{5}\oplus\overline{45}  \oplus \overline{70}}\,, \\
\mathbf{ {N} \> \otimes {\Sigma} } \>\> =& \mathbf{{1_S}}\oplus
\mathbf{{24_S}}  \oplus \mathbf{24_A} \oplus \mathbf{{75_S}}
  \oplus \mathbf{{126_A}}   \nonumber\\ 
&\oplus\mathbf{\overline{126_A}} \oplus \mathbf{{200_S}} \,,
\end{align}
where the subscripts $\scriptstyle\mathbf{S,\,A}$ in the last line
denote the symmetric or antisymmetric nature of the corresponding
representations.  We assume that all FN fields transform nontrivially
under $SU(5)$, and thus that no singlet exists and, for simplicity, we
restrict ourselves to representations with dimension less than 100,
which results in the following possibilities $\mathbf{R}=\mathbf{24}$,
$\mathbf{5}$, $\mathbf{45}$, $\mathbf{70}$.  

{\it Pointlike propagators:}\ Since the mass $M$ of
these fields is assumed to be larger than $\langle \Sigma_\pm\rangle
\sim \Lambda_{\text{GUT}}$, the contributions to the operators in
Eq.~\eqref{eq:4} can be evaluated by means of insertions of effective
pointlike propagators. As in \cite{Duque:2008ah} we denote the
contractions of two vectorlike fields in the representation
$\mathbf{R}$, $\overline{\mathbf{R}}$ as
\begin{align}
  \left[\mathbf{R}_{d e\ldots}^{a b c\ldots}\overline{\mathbf{R}}_{l m n\ldots}^{p
        q\ldots}\right]
  =-\frac{i}{M}\mathcal{S}_{d e \ldots l m n\ldots}^{a b c\ldots p q \ldots}\,,
\end{align}
where all the indices are $SU(5)$ indices, and $\mathcal{S}$ is the appropriate
group index structure. The structures $\mathcal{S}$ for
$\left[\fv^a\,\bfv_b\right]$,
$\left[\for^{ab}_c\,\bfor^n_{lm}\right]$ and
$\left[\se^{ab}_c\,\bse^n_{lm}\right]$
(and for several other
$SU(5)$ representations) can be found  in Appendix A of
\cite{Duque:2008ah}. In addition we need the following
contractions
\begin{align}
  \label{eq:10S}
i M \left[  \mathbf{24}^a_b\,\mathbf{24}^l_m  \right]_{\mathbf{S}}&= 
\left(\mathcal { S}_{\mathbf{S}}\right)^{a\,l}_{b\,m}=
\frac{5}{2}\left[\delta^a_m\, \delta_b^l +
\delta^a_l\, \delta_b^m\right]
- \delta^a_b\, \delta_m^l \,,  \\
  \label{eq:10A}
i M\left[\mathbf{24}^a_b\,\mathbf{24}^l_m  \right]_{\mathbf{A}}&=
\left(\mathcal{ S}_{\mathbf{A}}\right)^{a\,l}_{b\,m}= \frac{5}{2}
\left[\delta^a_m\, \delta_b^l -\delta^a_l\, \delta_b^m\right]\,.
\end{align}
These two expressions are obtained by imposing the traceless condition
for the adjoint $\left(\mathcal{ S}_{\mathbf{S,A}}\right)^{a\,l}_{a\,m}
=\left(\mathcal{ S}_{\mathbf{S,A}}\right)^{a\,l}_{b\,l}=0$ and the
normalization factor is fixed by the requirement that the (subtracted)
singlet piece $\delta^a_b\, \delta_m^l$ in Eq.~\eqref{eq:10S} provides
the proper singlet contraction, that is, by inserting the singlet in
the diagram of Fig.\ref{fig:1}(b) we require that the operator
$\left(\mathbf{\overline{5}}_a \mathbf{5}_{\phi_u}^a\right)
\cdot\left(\mathbf{N}_l^j \mathbf{\Sigma}_j^l\right)$ is obtained with
unit coefficient.  

{\it Vertices:}\  All the vertices we need involve $\mathbf{5}_{\phi_u}$ or the
adjoint $\bSigma$ with the external fermions $\bfv$ and $\mathbf{N}$,
or with the FN representations $\mathbf{R}$ in the internal lines.
The vertices have the general form $-i \lambda \mathcal{ V}$ where
$\lambda$ is universal for all vertices.  Including symmetry factors,
the relevant field contractions $ \mathcal{
  V}=\mathbf{R}\,\mathbf{5}_{\phi_u}\,\mathbf{R'} $ or $\mathcal
  {V}=\mathbf{R}\,\mathbf{\Sigma}\,\mathbf{R'}$, with
$\mathbf{R},\mathbf{R'}=\mathbf{5},\,\mathbf{24},\,\mathbf{45},\,\mathbf{70}$,
are:
\begin{eqnarray}
\label{eq:v5} &&
    \bfv_a \tf^a_b\fv^b \qquad  \!\!\!
\bfv_a \tf^c_b \for^{ba}_c \qquad  \!\!\!
\bfv_a \tf^c_b \se^{ba}_c 
\qquad \tf^a_c\tf^c_b
\left(\tf_{\mathbf{S,A}}\right)^b_a.
\\
\label{eq:v45} &&
\bfor_{ab}^c{\tf^\uparrow}^b_d\for^{da}_c \quad\    \quad
   \frac{1}{2}\, \bfor_{ab}^c{\tf^\downarrow}_c^d\for^{ba}_d  
 \\ \label{eq:v70} &&
\bse_{ab}^c{\tf^\uparrow}^b_d\se^{da}_c   \quad\   \quad
   \frac{1}{2}\, \bse_{ab}^c{\tf^\downarrow}_c^d\se^{ba}_d     
\quad\ \quad 
\bfor_{ab}^c\tf^b_d\se^{da}_c \,,  
\end{eqnarray}
where the vertices in the first line describe the couplings of the
external states ($\mathbf{\overline 5}$ and $\mathbf{N}$) with heavy
FN fields and flavons, while the last two lines involve only heavy FN
fields and flavons.  There are two inequivalent ways of contracting
the indices for the vertices involving the $\tf$ with pairs of $\for$
and $\se$ in the last two lines~\cite{Duque:2008ah}. They are distinguished in
Eqs.~(\ref{eq:v45}) and (\ref{eq:v70}) by an up ($\tf^\uparrow$) or
down ($\tf^\downarrow$) arrow-label.  As explained
in~\cite{Duque:2008ah}, this can be traced back to the fact that these
representations are contained twice in their tensor products with the
adjoint.

\begin{figure}[h!!]
  \centering

\includegraphics[scale=0.85]{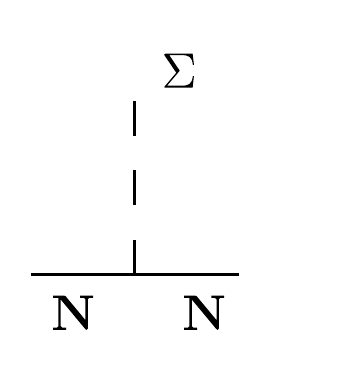} \hspace{1.0cm}
\includegraphics[scale=0.85]{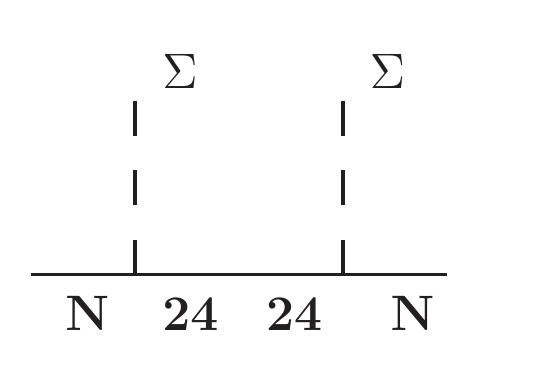} \hspace{1.0cm}
\includegraphics[scale=0.85]{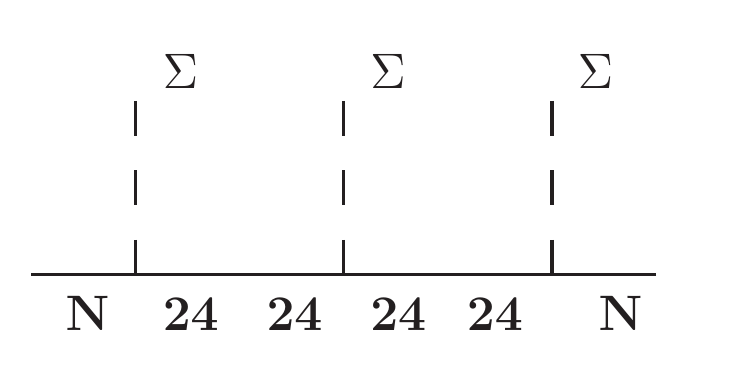} \\ [-5pt]

\leftline{\hspace{1.6cm}(a)\hspace{4.7cm}(b)\hspace{6.3cm}(c)}

\caption{
Diagrams contributing to  $M_R$ at different orders. The lowest order coefficient 
$r_1$ is obtained from diagram (a), $r_2$ from (b), and $r_3$ from (c).}
\label{fig:2}
\end{figure}

{\it Relevant multiplet components:}\ We write the $SU(5)\times
U(1)_H$ breaking vevs as
\begin{equation}
\label{eq:VEV}
\langle\mathbf{\Sigma_{\pm}}\rangle=
\frac{V}{\sqrt{60}}\times\operatorname{diag}(2,2,2,-3,-3)\,,
\end{equation}
where the factor $1/\sqrt{60}$ gives the usual normalization of the
$SU(5)$ generators,
$\operatorname{Tr}(\mathbf{R}^a\overline{\mathbf{R}^b})=(1/2)\delta^{ab}$,
and the coefficients of the left handed neutrino couplings to the
$SU(2)$ singlet $\nu\,\phi^0_u\,N_S$ and $SU(2)$ triplet
$\nu\,\phi^0_u\,N_T$ as well as the Majorana neutrinos mass terms
$N_{S,T}\,N_{S,T}$ are obtained by projecting the representations
$\mathbf{5}$, $\mathbf{5}_{\phi_u}$ and $\mathbf{N}$ onto the relevant
field components according to
\begin{eqnarray}
\label{eq:projections1}
\nu &=& - \overline{\mathbf{5}}_5 = - \delta_5^a \; \overline{\mathbf{5}}_a 
\\ 
\label{eq:projections2}
\phi^0_u &=&\mathbf{5}_{\phi_u}^5   =  \delta^5_b\; \mathbf{5}_{\phi_u}^b   
\\
\label{eq:projections3}
N_S &=& \frac{1}{\sqrt{60}}\;
\operatorname{diag}(2,\,2,\,2,\,-3,\,-3)\cdot \mathbf{N}_{24}\,.
\\
N_T &=& \frac{1}{\sqrt{60}}\;
\operatorname{diag}(0,\,0,\,0,\,\sqrt{15},\,-\sqrt{15})
\cdot \mathbf{N}_{3}\,.
\label{eq:projections4}
\end{eqnarray}
where the subscripts in $\mathbf{N}_{24}$ (singlet) and
$\mathbf{N}_{3}$ (neutral component of the triplet) refer to the
corresponding $SU(5)$ generators.  The assumption of a unique heavy
mass parameter $M$ for the FN fields and of universality of the
fundamental scalar-fermion couplings $\lambda$ yield a remarkable
level of predictivity.  In particular, for the vertices involving
$\mathbf{\Sigma_\pm}$ we can always reabsorb $\lambda V \to V$. This
leaves just an overall power of $\lambda$ common to all effective
Yukawa operators that involve one insertion of the Higgs multiplet
$\mathbf{5}_{\phi_u}$ (see the diagrams in Figs.~\ref{fig:1}) and no
$\lambda$ at all for the contributions to $M_R$, (see the diagrams in
Figs.~\ref{fig:2}).

The contributions to $Y_\nu$ and $M_R$ at different orders can be
computed using the vertices $\mathcal{V}$ given in
Eqs.~\eqref{eq:v5}-\eqref{eq:v70} and the relevant group structures
$\mathcal{S}$ in Eqs.~\eqref{eq:10S}, \eqref{eq:10A} and in Appendix A
of~\cite{Duque:2008ah}, that account for integrating out the heavy FN
fields. Additionally, the multiplets $\mathbf{\bar 5}$, $\mathbf{N}$,
and $\mathbf{5}_{\phi_u}$ in the external legs of the diagrams must be
projected on the relevant components according to
Eqs.~\eqref{eq:projections1}-\eqref{eq:projections4} and the flavons
$\mathbf{\Sigma_\pm}$ have to be projected onto the vacuum according to
Eq.~\eqref{eq:VEV}.

We have evaluated the $Y_\nu$ including the contributions up to
$\mathcal{O}(\epsilon^2)$ that are diagrammatically depicted in
Figs.~\ref{fig:1}: \ref{fig:1}(a) $\mathcal{O}(\epsilon^0)$;
\ref{fig:1}(b)--\ref{fig:1}(c) $\mathcal{O}(\epsilon^1)$;
\ref{fig:1}(d)--\ref{fig:1}(f) $\mathcal{O}(\epsilon^2)$.  $M_R$ has
been computed including contributions with three insertions 
of the flavons $\Sigma_\pm$ 
corresponding to the diagrams in Figs.~\ref{fig:2}: \ref{fig:2}(a)
$\mathcal{O}(\epsilon)$; \ref{fig:2}(b) $\mathcal{O}(\epsilon^2)$;
\ref{fig:2}(c) $\mathcal{O}(\epsilon^3)$.  At each specific order, 
the contributions to specific entries in $Y_\nu$ and $M_R$ 
can be written as 
\begin{eqnarray}
  \label{eq:orders}
  Y_\nu^{(i)} &=&  \lambda\,\alpha^{i+1}\,\epsilon^i \cdot  
\left(y^S_i +y^T_i\right)\,,
\\
M_R^{(i)}& =& V\,\alpha^{i+3}\,\epsilon^i 
\cdot \left(r^S_{i+1}+r^T_{i+1} \right)= 
M\,\alpha^{i+3}\,\epsilon^{i+1}
\cdot \left(r^S_{i+1}+r^T_{i+1} \right)\,,
 \end{eqnarray}
 where $\alpha= 1/\sqrt{60}$ is the normalization factor for
 $\mathbf{\Sigma}$ and for the $N_{S,T}$ in the adjoint, $V=M\epsilon$
 with $V$ defined in Eq.~(\ref{eq:VEV}), and $y_i^{S,T}$ and
 $r_{i+1}^{S,T}$ are the nontrivial group theoretical coefficients,
 that we have computed for $i=0,1,2$ and for the singlet $(S)$ and triplet
$(T)$ contributions to the seesaw Lagrangian. 
The corresponding
 results for $y_i^{S,T}$ are given in Table~\ref{tab:1} (where we have
 followed the notation of~\cite{Duque:2008ah}), while the results for
 $r_{i+1}^{S,T}$ are given in Table~\ref{tab:2}.

%

\renewcommand{\arraystretch}{1.28}
\begin{table}[t!]
  \centering
  \begin{tabular}{c|c|c|c|c|c|c|c|c|c}
    $\mathbf{\bar 5}_1$ & $\mathbf{\bar 5}_2$ & $\mathbf{\bar 5}_3$ 
& $\mathbf{10}_1$ & $\mathbf{10}_2$ & $\mathbf{10}_3$ &
    $\mathbf{5}_{\phi_u}=-\mathbf{5}_{\phi_d}$&$N_1$&$N_2$ &$N_3$\\\hline
    $-\frac{29}{10}$& -$\frac{9}{10}$ & $\frac{31}{10}$&
$\frac{13}{10}$&$\frac{3}{10}$&$-\frac{7}{10}$&$\frac{7}{5}$&
$\frac{5}{2}$&$-\frac{1}{2}$&$-\frac{11}{2}$\\
  \end{tabular}
  \caption{$F$--charges obtained with $n=-6$ in Eq.~\eqref{eq:2}, 
and $m_1=2$, $m_2=-1$, and $m_3=-6$ in Eq.~\eqref{eq:3}.} 
  \label{tab:3}
\end{table}
\renewcommand{\arraystretch}{1}

We have searched for all possible charge assignments with absolute
values of the $F$ charges smaller than 10, and we have examined the
resulting neutrino mass matrices.  We have found some promising
possibilities.  If we choose, for example, in Eqs.~(\ref{eq:5}) and
(\ref{eq:6}), $n=-6$ and $m_1=2$, $m_2=-1$, $m_3=-6$, we obtain the
$F$--charges shown in Table~\ref{tab:3}, which can be obtained from
the set given in Eq.~\eqref{eq:1} through the redefinitions
Eqs.~\eqref{eq:shift} with $a_{-6}=21/10$.

According to Eq.~\eqref{eq:5}, this set of $F$--charges 
gives the following orders of magnitude for $Y_\nu$:
\begin{align}
  \label{eq:12}
    Y_{\nu}\sim&\lambda
  \begin{pmatrix}
\epsilon   & \epsilon^2    & \epsilon^7 \\
\epsilon^3 & 1           & \epsilon^5 \\
\epsilon^7 &  \epsilon^4 & \epsilon \\
  \end{pmatrix}.
\end{align}
Neglecting terms of $\mathcal{O}(\epsilon^4)$ and higher, 
including the coefficients $y_i^{S,T}$ and the 
appropriate powers of the normalization factor $\alpha$, this reads:
\begin{align}
 Y_\nu^{S,T}\,\approx\,&\lambda\, \alpha\,  \begin{pmatrix}
y_1(\alpha\epsilon)  & y_2(\alpha\epsilon)^2 & 0 \\
y_3(\alpha\epsilon)^3 & y_0  & 0 \\
0 & 0  & y_1(\alpha\epsilon) \\
  \end{pmatrix}^{S,T}.
\end{align}
where the superscript $^{S,T}$ outside the matrix is a shorthand for
$y_i^{S,T}$ inside the matrix.  Similarly, according to
Eq.~\eqref{eq:6} and \eqref{eq:8} we have for the entries in $M_R$ the
following orders of magnitude:
\begin{align}
  \label{eq:13}
M_R\,\sim\, & 
V\,   
  \begin{pmatrix}
\epsilon^4 & \epsilon^1 & \epsilon^2\\
\epsilon^1 & \epsilon^0 &\epsilon^5   \\
\epsilon^2 & \epsilon^5 &\epsilon^{10}   \\
  \end{pmatrix}.
\end{align}
Neglecting terms of $\mathcal{O}(\epsilon^4)$ and higher, and 
taking into account the coefficients $r_i^{S,T}$ and $\alpha$, we obtain
\begin{align}
\label{eq:14}
 M_R^{S,T} \approx\, & \, V \, \alpha^{3}\, 
\begin{pmatrix}
0   & r_2 (\alpha\epsilon) & r_3(\alpha\epsilon)^2\\
r_2(\alpha\epsilon) & r_1 &0   \\
r_3(\alpha\epsilon)^2 & 0 &0   \\
  \end{pmatrix}^{S,T} \,.
\end{align}
According Eq.~\eqref{eq:7}, the resulting light neutrino mass matrix  then is 
\begin{align}
\label{eq:Mnuappx}
M_\nu\approx& - \frac{v^2\sin^2\beta}{\alpha V}
  \,\lambda^{2}\sum_{S,T} \left[ \frac{1}{r_1\,r_3}\, \begin{pmatrix}
0    &  0 & y_1^2r_1 \\
0    & \phantom{\Big|}\!\!y_0^2r_3 & -y_0y_1r_2 \\ 
y_1^2 r_1 & -y_0y_1r_2 & \frac{1}{r_3}\, y_1^2 r_2^2
  \end{pmatrix}\right],
\end{align}
where we have neglected in each entry corrections of $\mathcal
  {O}(\alpha\epsilon)^2$ and higher, and we have suppressed the
subscripts $^{S,T}$ not to clutter the expression. It is remarkable
that at leading order the structure of the light neutrino mass matrix
remains determined only in terms of the group theoretical coefficients
$y_i^{S,T}$ and $r_i^{S,T}$, and in particular it does not depend on the
hierarchical parameter $\epsilon$. Let us also note that this matrix
corresponds to the two zero--texture type of neutrino mass matrix
discussed in~\cite{Mohanta:2006xd}.  As regards the scale $\alpha V$
appearing in the denominator of Eq.~\eqref{eq:Mnuappx}, it can be
directly related with the unification scale, defined as the mass scale
of the leptoquarks gauge fields $M_X=M_Y$~\cite{Bailin:1986wt}:
\begin{align}
\label{eq:Lgut}
  \Lambda_{GUT}=M_X
=5\,g_5\,\alpha\, V\,,
\end{align}
where $g_5\approx 0.7$ is the unified gauge coupling at
$\Lambda_{GUT}\simeq10^{16}$.


It is remarkable to note that both $Y_\nu$ and $M_R$ are hierarchical,
with the first one having a hierarchy between its eigenvalues of
$\mathcal {O}(\alpha\epsilon)$ and the second one of $\mathcal
  {O}(\alpha\epsilon^2)$.  The light neutrino mass matrix computed
naively (and erroneously, see Section~\ref{sec:signs}) from the
effective seesaw operator using only the charges of the
$\overline{\mathbf{5}}_I$ multiplets, would also be hierarchical.
However, the resulting $M_\nu$ is not hierarchical, and in fact at
leading order it does not depend at all on $\epsilon$ but only on the
group theoretical coefficients $y_i^{S,T}$ and $r_i^{S,T}$.  It is precisely the
presence of $F$ charges of both signs for the fields and for the two
flavons that yields the possibility of obtaining non-hierarchical
neutrino masses and large mixing angles, although the whole scenario
is defined at the fundamental level in terms of a small hierarchical
parameter $\epsilon$.

Let us comment at this point that, as it is discussed
in~\cite{Duque:2008ah}, corrections from sets of higher order diagrams
to the various entries in $Y_\nu$ and $M_R$ can generically be quite
sizable, although suppressed by higher powers of $\epsilon$. This is
because at higher orders the number of diagrams contributing to the
various operators proliferate, and the individual group theoretical
coefficients also become generically much larger, as can be seen in
Tables~\ref{tab:1} and~\ref{tab:2}.  By direct evaluation of higher
orders corrections, the related effects were estimated
in~\cite{Duque:2008ah} to be typically of a relative order $\sim 20\%
- 30\%$. To take into account the possible effects of these
corrections, we allow for a $\sim 25\% $ uncertainty in  the 
final numerical results.

\section{Numerical analysis}
\label{sec:numerical}


\begin{table}[t!]
  \centering
\begin{tabular}{|l|r|r||l|r|r|}\hline
\phantom{$\Big|$}$\epsilon^1$ & $y_1^S$& $y_1^T$&\ $\epsilon^2$ &$y_2^S$&$y_2^T$ \  \\
\hline\phantom{$\Big|$}
$\!\![\mathbf{5}^{\phi_{u}}\Sigma]$& & &$[\mathbf{5}^{\phi_{u}}\Sigma\Sigma]$&$ $&$$ \ \\  [3pt]
$O(\epsilon;\mathbf{24_S})$&$-15$&$-15\sqrt{15}$&$O(\epsilon^{2};\mathbf{24_S},\mathbf{24_S})$&$-75$&$-225\sqrt{15}$\\[2pt]
$O(\epsilon;\mathbf{24_A})$&$\ 0$&$0$&$O(\epsilon^{2};\mathbf{24_A},\mathbf{24_S})$& $0$&$0$\\ [2pt]
&&&$O(\epsilon^{2};\mathbf{24_S},\mathbf{24_A})$& $0$&$0$\\ [2pt]
&&&$O(\epsilon^{2};\mathbf{24_A},\mathbf{24_A})$& $0$&$0$\\  [3pt]
\hline \phantom{$\Big|$}
$[\Sigma\mathbf{5}^{\phi_{u}}]$& & &$[\Sigma\mathbf{5}^{\phi_{u}}\Sigma]$&$ $&$$ \\ [3pt]
$O(\epsilon;\mathbf{5})$&$\ -9$&$-3\sqrt{15}$&$O(\epsilon^{2};\mathbf{5},\mathbf{24_S})$&$-45$&$-45\sqrt{15}$\\  [2pt]
&&&$O(\epsilon^{2};\mathbf{5},\mathbf{24_A})$& $\ \ 0$&$0$\\  [2pt]
$O(\epsilon;\mathbf{45})$&$\  75$&$-15\sqrt{15}$&$O(\epsilon^{2};\mathbf{45},\mathbf{24_S})$&$300$&$-180\sqrt{15}$\\ [2pt]
&&&$O(\epsilon^{2};\mathbf{45},\mathbf{24_A})$&$\ \ 0$&$0$\\ [2pt]
$O(\epsilon;\mathbf{70})$&$-225$&$-15\sqrt{15}$&$O(\epsilon^{2};\mathbf{70},\mathbf{24_S})$&$-900$&$-180\sqrt{15}$\\ [2pt]
&&&$O(\epsilon^{2};\mathbf{70},\mathbf{24_A})$& $\ \ 0$&$0$\\ [2pt]
\cline{4-6}& &\phantom{$\Big|$}
&$[\Sigma\Sigma\mathbf{5}^{\phi_{u}}]$&$ $&$$                    \\ [3pt]
&&&$O(\epsilon^{2};\mathbf{5},\mathbf{5})$& $-27$&$-9\sqrt{15}$             \\ [2pt]
&&&$O(\epsilon^{2};\mathbf{5},\mathbf{45})$& $225$ &$-45\sqrt{15}$            \\ [2pt]
&&&$O(\epsilon^{2};\mathbf{5},\mathbf{70})$& $-675$&$-45\sqrt{15}$           \\ [2pt]
&&&$O(\epsilon^{2};\mathbf{45},\mathbf{5})$& $225$&$75\sqrt{15}$             \\ [2pt]
&&&$O^{\uparrow}(\epsilon^{2};\mathbf{45},\mathbf{45})$& $1425$&$-285\sqrt{15}$ \\ [2pt]
&&&$O^{\downarrow}(\epsilon^{2};\mathbf{45},\mathbf{45})$&$525$&$-105\sqrt{15}$\\ [2pt]
&&&$O(\epsilon^{2};\mathbf{45},\mathbf{70})$& $1125$&$75\sqrt{15}$           \\ [2pt]
&&&$O(\epsilon^{2};\mathbf{70},\mathbf{5})$& $-675$&$-225\sqrt{15}$           \\ [2pt]
&&&$O(\epsilon^{2};\mathbf{70},\mathbf{45})$& $-1125$&$225\sqrt{15}$          \\ [2pt]
&&&$O^\uparrow(\epsilon^{2};\mathbf{70},\mathbf{70})$& $-4725$&$-315\sqrt{15}$ \\ [2pt]
&&&$O^\downarrow(\epsilon^{2};\mathbf{70},\mathbf{70})$& $-675$&$-45\sqrt{15}$\\ [3pt]
\hline\phantom{$\Big|$}
$\Sigma_{\mathbf{R}}O(\epsilon; \mathbf{R})$&$-174$&$-48\sqrt{15}$&$\Sigma_{\mathbf{R}} O(\epsilon^{2}; \mathbf{R})$&$-5097$&$-1329\sqrt{15}$\\ [3pt]
\hline
\end{tabular}
\caption{Operators contributing to  ${Y}_\nu=\sum_i {Y}^{(i)}_\nu$ 
  at $\mathcal{ O}(\epsilon)$ and  $\mathcal{ O}(\epsilon^2)$ and  values of 
  the corresponding coefficients 
  $y_i =Y^{(i)}_\nu/\left(\lambda\,\alpha^{i+1}\, \epsilon^i\right)$
  for the singlet $(S)$ and triplet $(T)$ components.  
  The value of the  $\mathcal{ O}(1)$ coefficients are 
  $y_0^S=3$ and $y_0^T=\sqrt{15}$. An example of calculating an entry in the table is given in Appendix \ref{aped.C}
}
\label{tab:1}
\end{table}
%

%
\begin{table}[t!] 
  \centering
\begin{tabular}{|l|c||l|c||l|c|}\hline
\phantom{$\Big|$}$\epsilon^1$&$(r_1^S,r_1^T)$
& $\epsilon^2$ & $(r_2^S,r_2^T)$
& $\epsilon^3$ & $(r_3^S,r_3^T)$
\\\hline
\phantom{$\Big|$}
$\!\![\Sigma]$&&$[\Sigma\Sigma]$&& $[\Sigma\Sigma\Sigma]$&$$ \\  [6pt]
\phantom{$\Big|$}$O(\epsilon;\mathbf{ 24})$ & $(-30,-90)$
&$O(\epsilon^{2};\mathbf{ 24_S})$& $(-150,-1350)$
&$O(\epsilon^{3};\mathbf{24_S},\mathbf{24_S})$& $(-750,-20250)$ \\  [4pt]
&&$O(\epsilon^{2};\mathbf{ 24_A})$& $(0,0)$
&$O(\epsilon^{3};\mathbf{24_A},\mathbf{24_S})$& $(0,0)$\\ [4pt]
&& &&$O(\epsilon^{3};\mathbf{24_S},\mathbf{24_A})$& $(0,0)$\\ [4pt]
&& &&$O(\epsilon^{3};\mathbf{24_A},\mathbf{24_A})$& $(0,0)$\\ [4pt]
\hline \phantom{$\Big|$}
$\Sigma_{\mathbf{R}} O(\epsilon; \mathbf{R})$&$(-30,-90)$
&$\Sigma_{\mathbf{R}} O(\epsilon^{2}; \mathbf{R})$&$(-150,-1350)$
&$\Sigma_{\mathbf{R}} O(\epsilon^{3}; \mathbf{R})$&$(-750,-20250)$ \\ \hline
\end{tabular}
\caption{
  Operators contributing to  ${M}_R=\sum_i M^{i}_R$  at 
  $\mathcal{ O}(\epsilon)$, $\mathcal{O}(\epsilon^2)$  
  and  $\mathcal{ O}(\epsilon^3)$, and values of the corresponding coefficients 
  $r_i =M_R^{(i-1)}/\left(V\alpha^{i+2}\,\epsilon^{i-1}\right)$ for the singlet $(S)$ and triplet $(T)$ components.  
}
\label{tab:2}
\end{table}

Allowing for all the contributions listed in Table~\ref{tab:1}, the
resulting coefficient at $\mathcal{ O}(\epsilon)$ for $Y_\nu^S$ would be
$y_1^S=-174$ that is too large to reproduce the neutrino oscillation
data.  We will then assume that only some contributions are present.
This is easily achieved by assuming that no FN fields exist in the
representations $\mathbf{70}_{39/10}$, $\mathbf{70}_{-41/10}$,
$\mathbf{45}_{39/10}$ and $\mathbf{45}_{-41/10}$, and this results in
much smaller coefficients $y_1^S=-24$ and $y_1^T=-18\sqrt{15}$ that
are determined by the $y_1^{S,T}$ entries in the first and third lines
in Table~\ref{tab:1}, and that are the one we will use henceforth.
(The absence of these representations also implies that several
contributions to the higher order coefficient $y_2^{S,T}$ are absent,
which yields much smaller values $y_2^{S,T}\sim 10^2$ instead than
$\sim 10^3$, see Table~\ref{tab:1}.  In any case, since at leading
order $M_\nu$ Eq.~(\ref{eq:Mnuappx}) does not depend on $y_2^{S,T}$,
this only affects the higher order corrections.) As regards the
contributions to $M_R$, they arise only from insertions of the
$\mathbf{24}$, and thus they are not affected by the absence of
$\mathbf{70}$ and $\mathbf{45}$.

By using in
Eq.~(\ref{eq:Mnuappx}) $(y_0^S,y_0^T)=(3,\sqrt{15})$, 
$(y_1^S,y_1^T)=(-24,-18\sqrt{15})$ 
and the values of $(r_i^S,r_i^T)$ given  in
Table~\ref{tab:2}, we  obtain
\begin{align}
\label{eq:15}
 M_\nu \;\approx\; & - 5\, 
\left(\lambda\,\sin\beta\right)^2
\frac{g_5\, v^2}{\Lambda_{\text{GUT}}}
   \, 
\begin{pmatrix}
0    & 0 & -1.0 \\
0    & -0.47 & -0.68 \\
-1.0 & -0.68 & -1.0
  \end{pmatrix}.
\end{align}
With $v=175\,$GeV and $\Lambda_{\text{GUT}}\approx 10^{16}\,$GeV
 the numerical value of the
prefactor is $ 
\approx 0.008\,(\sin\beta\,\lambda)^2\,$eV.  For $\tan\beta\approx10$
($\tan\beta\approx1$) the atmospheric mass scale $\approx 0.05\,$eV
can then be reproduced for acceptable values of the coupling
$\lambda\sim 1.9\;(2.7)\,$.

Our model is based on the successful model for the $d$-quark and
leptons masses discussed in Ref.~\cite{Duque:2008ah}, and we have
checked that the absence of the representations that we have forbidden
here do not affect the results of this previous study.  In particular,
by using the coefficients calculated in Ref.~\cite{Duque:2008ah} we
have for the matrix of the charged leptons Yukawa couplings
\begin{align}
  Y^e\simeq
  \begin{pmatrix}
  \epsilon^4 & \epsilon^5 & \epsilon^4 \\
 -2.9\epsilon^3 & 3.8\epsilon^2 
& 10.2\epsilon^3 \\
 -7.6\epsilon^3 & 9.2\epsilon^2 & 2.3\epsilon 
  \end{pmatrix}\,.
\end{align}
To compute neutrino mixing matrix
$U_{PMNS}=U_\nu\left(V_L^e\right)^\dagger$, besides the matrix $U_\nu$
that diagonalizes $M_\nu$ in Eq.~\eqref{eq:15}, we also need $V_L^e$
that diagonalizes the left-handed product ${Y^e}{Y^e}^\dagger$. We
obtain
\begin{equation}
  \label{eq:Ue}
V_L^e \sim 
  \begin{pmatrix}
1.     & 10^{-5} &10^{-5}  \\
10^{-5} & -1 & 0.02 \\
 10^{-5} & 0.02 & 1
  \end{pmatrix}.
\end{equation}
that is approximately diagonal, and thus $U_{PMNS}\approx U_\nu$.
Allowing for a $\sim 25\%$ numerical uncertainty in the entries of the
matrix in Eq.~\eqref{eq:15}, we find that it is possible to fit the
neutrino oscillation data, with the exception of
$\sin^2\theta_{12}\sim 0.5$ for which a particularly large corrections
is needed. Finally, the mass of the lightest heavy singlet and triplet
neutrino states can be obtained from Eq.~\eqref{eq:14} and are
\begin{align}
   & M_1^S\ \approx\ 5\times 10^{11}\, \text{GeV}\,,\\ 
&  M_1^T\approx 1.5\times 10^{13}\, \text{GeV}\,.
\end{align}
In particular the mass of the singlet Majorana neutrino is of the
right order of magnitude to allow for thermal
leptogenesis~\cite{Davidson:2008bu}.


\chapter{Conclusions}\label{sec:conlusions}

In this thesis we have obtained a supersymmetric $R$-parity breaking model with $B$
violation by considering the most general supersymmetric standard
model allowed by gauge invariance and extending it with a
SFH $U(1)_H$ symmetry. 
The generated effective theory at low energy has only the particle
content of the SSM.
After imposing existing constraints in both single and quadratic
$R$-parity violating (RPV) couplings, only one precise
hierarchy remains depending on a global suppression factor
$\theta^{n_{\lambda''}}$ ($n_{\lambda}''>1$) with $\lambda''_{323}$ as
the dominant coupling and very suppressed couplings for the first two
generations.
Additional suppression is required in order to obtain Dirac neutrino
masses in the model, and only solutions with $n_{\lambda}''\ge6$
remain allowed.    
In this way, the resulting RPV and $B$-violating model also explaining
neutrino masses is powerful enough to satisfy all the existing constraints
on RPV. 
In particular, the $U(1)_H$ symmetry also ensures that dimension-five
$L$-violating operators are sufficiently suppressed so that
the decay of the proton is above the experimental limits.
The resulting underlying theory for the RPV operators is quite
similar to that obtained after imposing the MFV hypothesis on a general RPV model (at least until couplings of
order $\theta^{n_{\lambda''}+3}$), and therefore the predictions of
both models are the same at the LHC.
The phenomenology at colliders depends strongly on the nature and decay length of
the LSP. 
Specific searches at the LHC for the RPV with $B$ violation have reported
restrictions only in the case of prompt decays of the gravitino when
it is the LSP.  
Several analyses of CMS and ATLAS involving leptons have been
reanalyzed to constrain the gluino as a function of the stop mass
(see Ref.~\cite{Berger:2013sir} and references therein) within a special spectrum
guaranteeing that $\operatorname{BR}(\tilde{g}\to
\tilde{t}\,\bar{t})=1$ and with prompt decays of the corresponding
LSP stop. 
In both cases bounds in the gluino mass around $600\text{ GeV}$ have
been obtained. Therefore, the parameter space of the RPV/SFH scenario
(or the RPV/MFV one) have still plenty of room to accommodate a low-energy 
supersymmetric spectrum.
There is a number of open issues that could be more easily studied
within this realistic and predictive framework,  for example, the
constraints on the couplings from low-energy observables and indirect
dark matter experiments or the restrictions in the parameter space
from other collider signatures like the displaced vertices searches
already implemented by ATLAS~\cite{Aad:2012zx} and
CMS~\cite{Chatrchyan:2012jna}.

Also we have worked in a $SU(5)\times U(1)_H$ model for charged fermion
masses studied in Ref.~\cite{Duque:2008ah} to include neutrino masses.
This has been done by means of an appropriate redefinition of the
$U(1)_H$ charges that, while it leaves unchanged the Yukawa matrices
for the charged fermions, it also forbids at all orders $\Delta B\neq
0$ and $\Delta L=1$ operators, while allowing for $\Delta L=2$
Majorana mass terms. Thus, $R$-parity is enforced as an exact
symmetry, but at the same time the seesaw mechanism can be embedded
within the model.  Our construction is severely constrained by two
theoretical requirements. First, the $SU(5)$ GUT implies that the $F$
charges of the lepton doublets and $d$-quarks singlets, as well as the
$F$ charges of the quark-doublets and lepton singlets are the same,
reducing drastically the freedom one has in the SM. Second, we have
assumed universality of all the fundamental scalar-fermion couplings,
which basically implies that the model has only one free parameter,
that is the ratio between the $U(1)_H$ breaking vevs and the messenger
scale $M$. In spite of these serious restrictions, we have shown that
by assigning the $U(1)_H$ breaking flavons to the adjoint of $SU(5)$,
computable group theoretical coefficients arise that, at leading
order, determine the structure of the neutrino mass matrix in a
parameter independent way. This structure yields a reasonable first
approximation to the measured neutrino parameters.  However, higher
order corrections can be large, and should be taken into account for a
more precise quantitative comparison with observations.  In our model,
hierarchical heavy Majorana neutrinos naturally coexist with
non-hierarchical light neutrinos, the atmospheric scale is easily
reproduced for natural values of the parameters, and the mass of the
lightest heavy neutral states, that lies about five order of
magnitude below the GUT scale, is optimal for leptogenesis.  

At the quantitative level, the predictivity of the model clearly 
relies on the assumption of universality of the Yukawa couplings. We
have not put forth any speculation concerning the fundamental physics
that might underlie such a strong assumption, but have merely adopted
it as a working hypothesis to highlight how a theory of calculable
`order one coefficients' might actually emerge in GUT models relying
just on a generalized FN mechanism.  Needless to say, by relaxing the
assumption of universality by a certain quantitative amount, all the
predictions would acquire a correspondent numerical uncertainty,
although the main qualitative features of the model will remain
unchanged.

\chapter{Agradecimientos}

Son insuficientes los poco (cuarenta, cincuenta, sesenta, ...) a\~nos que me han sido otorgados para agradecerle al profesor Diego Restrepo por su colaboraci\'on, paciencia y amabilidad durante mi camino en el aprendizaje de la f\'isica de part\'iculas. Gracias Diego.\\\\
Deseo tambi\'en expresar mi gratitud y colaboraci\'on a los dem\'as integrantes del Grupo de Fenomenolog\'ia de las Interacciones Fundamentales (GFIF). \\\\
Expreso mis agradecimientos al profesor Jos\'e Valle y a su equipo de investigaci\'on, en la Universidad de Valencia, por su amable acogida durante mi estancia.\\ \\
Le quiero agradecer al maestro Jaime Chica por ense\~narme la pasi\'on desinteresada que se debe tener hacia la ciencia.\\\\
Le expreso mis agradecimientos a mi esposa, Luz Amparo, por aguantar mi constante silencio durante estos a\~nos.\\\\

Finalmente, le doy las gracias a la Universidad de Antioquia.

\appendix
\chapter{Appendix}\label{apen1}


\section{$H$-charges of dimension-5 operators}\label{aped.B}
The horizontal charges for the dimension-5 operators that violate only $B$ are given by
\begin{align}\label{eq:D5BNV1}\nonumber
H\left[(\kappa_3)_{1jk}\widehat{Q}_1\widehat{Q}_j\widehat{Q}_k\widehat{H}_d\right]=&
A_3+(2x+4-n_{\lambda''})\mathbf{1_3},\\\nonumber
H\left[(\kappa_3)_{2jk}\widehat{Q}_2\widehat{Q}_j\widehat{Q}_k\widehat{H}_d\right]=&
A_3+(2x+3-n_{\lambda''})\mathbf{1_3},\\
H\left[(\kappa_3)_{3jk}\widehat{Q}_3\widehat{Q}_j\widehat{Q}_k\widehat{H}_d\right]=&
A_3+(2x+1-n_{\lambda''})\mathbf{1_3},
\end{align}
\begin{align}\label{eq:D5BNV2}\nonumber
H\left[(\kappa_{10})_{ij1}\widehat{Q}_i\widehat{Q}_j\widehat{d}^*_1\right]=&
A_3+(x-n_{\lambda''})\mathbf{1_3},\\\nonumber
H\left[(\kappa_{10})_{ij2}\widehat{Q}_i\widehat{Q}_j\widehat{d}^*_2\right]=&
A_3+(x+1-n_{\lambda''})\mathbf{1_3},\\
H\left[(\kappa_{10})_{ij2}\widehat{Q}_i\widehat{Q}_j\widehat{d}^*_3\right]=&H\left[(\kappa_{10})_{ij3}\widehat{Q}_i\widehat{Q}_j\widehat{d}^*_2\right].
\end{align}
For the lepton and baryon-number violating operators we have that
\begin{align}\label{eq:D5BLNV1}\nonumber
 H\left[(\kappa_{1})_{1jkl}\widehat{Q}_1\widehat{Q}_j\widehat{Q}_k\widehat{L}_l\right]=&
A_1+(5+2x+n_l-n_{\lambda''})\mathbf{1_3},\\\nonumber
 H\left[(\kappa_{1})_{2jkl}\widehat{Q}_2\widehat{Q}_j\widehat{Q}_k\widehat{L}_l\right]=&
A_1+(4+2x+n_l-n_{\lambda''})\mathbf{1_3},\\
 H\left[(\kappa_{1})_{3jkl}\widehat{Q}_3\widehat{Q}_j\widehat{Q}_k\widehat{L}_l\right]=&
A_1+(2+2x+n_l-n_{\lambda''})\mathbf{1_3},
\end{align}
\begin{align}\label{eq:D5BLNV2}\nonumber
H\left[(\kappa_{2})_{ij11}\widehat{u}_i\widehat{u}_j\widehat{d}_1\widehat{e}_1\right]=&
A_2+(6-n_1+n_{\lambda''})\mathbf{1_3},\\\nonumber
H\left[(\kappa_{2})_{ij21}\widehat{u}_i\widehat{u}_j\widehat{d}_2\widehat{e}_1\right]=&
A_2+(5-n_1+n_{\lambda''})\mathbf{1_3},\\\nonumber
H\left[(\kappa_{2})_{ij31}\widehat{u}_i\widehat{u}_j\widehat{d}_3\widehat{e}_1\right]=&
H\left[(\kappa_{2})_{ij21}\widehat{u}_i\widehat{u}_j\widehat{d}_2\widehat{e}_1\right],\\\nonumber
H\left[(\kappa_{2})_{ij12}\widehat{u}_i\widehat{u}_j\widehat{d}_1\widehat{e}_2\right]=&
A_2+(3-n_2+n_{\lambda''})\mathbf{1_3},\\\nonumber
H\left[(\kappa_{2})_{ij22}\widehat{u}_i\widehat{u}_j\widehat{d}_2\widehat{e}_2\right]=&
A_2+(2-n_2+n_{\lambda''})\mathbf{1_3},\\\nonumber
H\left[(\kappa_{2})_{ij32}\widehat{u}_i\widehat{u}_j\widehat{d}_2\widehat{e}_2\right]=&H\left[(\kappa_{2})_{ij22}\widehat{u}_i\widehat{u}_j\widehat{d}_3\widehat{e}_2\right]\\\nonumber
H\left[(\kappa_{2})_{ij13}\widehat{u}_i\widehat{u}_j\widehat{d}_1\widehat{e}_3\right]=&
A_2+(1-n_3+n_{\lambda''})\mathbf{1_3},\\\nonumber
H\left[(\kappa_{2})_{ij23}\widehat{u}_i\widehat{u}_j\widehat{d}_2\widehat{e}_3\right]=&
A_2+(-n_3+n_{\lambda''})\mathbf{1_3},\\
H\left[(\kappa_{2})_{ij33}\widehat{u}_i\widehat{u}_j\widehat{d}_2\widehat{e}_3\right]=&H\left[(\kappa_{2})_{ij23}\widehat{u}_i\widehat{u}_j\widehat{d}_3\widehat{e}_3\right].
\end{align}
Finally, for the lepton-number violating terms we have found
\begin{align}\label{eq:D5LNV}\nonumber
H\left[(\kappa_{4})_{ij1}\widehat{Q}_i\widehat{H}_d\widehat{u}_j\widehat{e}_1\right]=&
A_4+(5-n_1+x)\mathbf{1_3},\\\nonumber
H\left[(\kappa_{4})_{ij2}\widehat{Q}_i\widehat{H}_d\widehat{u}_j\widehat{e}_2\right]=&
A_4+(2-n_2+x)\mathbf{1_3},\\\nonumber
H\left[(\kappa_{4})_{ij3}\widehat{Q}_i\widehat{H}_d\widehat{u}_j\widehat{e}_3\right]=&
A_4+(-n_3+x)\mathbf{1_3},\\\nonumber
H\left[(\kappa_{5})_{ij}\widehat{L}_i\widehat{H}_u\widehat{L}_j\widehat{H}_u\right]=&
\left(
\begin{array}{ccc}
 2n_1 & n_1+n_2 & n_1+n_3 \\
 n_1+n_2 & 2n_2 & n_2+n_3 \\
 n_1+n_3 & n_2+n_3 & 2n_3
\end{array}
\right),\\\nonumber
H\left[(\kappa_{6})_{i}\widehat{L}_i\widehat{H}_u\widehat{H}_d\widehat{H}_u\right]=&-1+n_i,\\\nonumber
H\left[(\kappa_{7})_{ij1}\widehat{u}_i\widehat{d}^*_j\widehat{e}_1\right]=&
A_7+(4-n_1)\mathbf{1_3},\\\nonumber
H\left[(\kappa_{7})_{ij2}\widehat{u}_i\widehat{d}^*_j\widehat{e}_2\right]=&
A_7+(1-n_2)\mathbf{1_3},\\\nonumber
H\left[(\kappa_{7})_{ij3}\widehat{u}_i\widehat{d}^*_j\widehat{e}_3\right]=&
A_7+(-1-n_3)\mathbf{1_3},\\\nonumber
H\left[(\kappa_{8})_{1}\widehat{H}_u^*\widehat{H}_d\widehat{e}_1\right]=&5-n_1+x,\\\nonumber
H\left[(\kappa_{8})_{2}\widehat{H}_u^*\widehat{H}_d\widehat{e}_2\right]=&2-n_2+x,\\\nonumber
H\left[(\kappa_{8})_{3}\widehat{H}_u^*\widehat{H}_d\widehat{e}_3\right]=&-n_3+x,\\
H\left[(\kappa_{9})_{i1k}\widehat{Q}_i\widehat{L}^*_j\widehat{u}_k\right]=&
A_9+(-n_j)\mathbf{1_3}.
\end{align}
In the above expressions we have defined 
\begin{align}
A_1&=A_3=\left(
\begin{array}{ccc}
 6 & 5 & 3 \\
 5 & 4 & 2 \\
 3 & 2 & 0
\end{array}
\right), \, 
A_2=\left(
\begin{array}{ccc}
 10 & 7 & 5 \\
 7 & 4 & 2 \\
 5 & 2 & 0
\end{array}
\right),\,
A_4=A_9=\left(
\begin{array}{ccc}
 8 & 5 & 3 \\
 7 & 4 & 2 \\
 5 & 2 & 0
\end{array}
\right),\,
A_7=\left(
\begin{array}{ccc}
 5 & 6 & 6 \\
 2 & 3 & 3 \\
 0 & 1 & 1
\end{array}
\right).
\end{align}

\section{Operator contributing to  ${Y}_\nu=\sum_i {Y}^{(i)}_\nu$ 
  at $\mathcal{ O}(\epsilon)$}\label{aped.C}

In this Appendix we calculate, as an example, one of the contributions of the effective Lagrangian associated with the representations of $SU(5)$  given in the Table \ref{tab:1}.
We write the $SU(5)\times U(1)_F$ breaking vevs as
\begin{equation}
\label{rep1}
\langle\mathbf{\Sigma_{\pm}}\rangle=
\frac{V}{\sqrt{60}}\times\operatorname{diag}(2,2,2,-3,-3)\,,
\end{equation}
where the factor $\alpha=1/\sqrt{60}$ gives the usual normalization of the
$SU(5)$ generators.
\begin{eqnarray}
\label{rep2}
N_S &=& \frac{1}{\sqrt{60}}\;
\operatorname{diag}(2,\,2,\,2,\,-3,\,-3)\cdot \mathbf{N}_{24}\,.
\end{eqnarray}
where the subscript in $\mathbf{N}_{24}$ (singlet) and refer to the
corresponding $SU(5)$ generators.
we need the following contraction
\begin{align}
  \label{rep3}
i M \left[  \mathbf{24}^a_b\,\mathbf{24}^l_m  \right]_{\mathbf{S}}&= \left(\mathcal { S}_{\mathbf{S}}\right)^{a\,l}_{b\,m}= \frac{5}{2}\left[\delta^a_m\,\delta_b^l + 
\delta^a_l\, \delta_b^m\right]- \delta^a_b\, \delta_m^l \,,  
\end{align}
Using the Eqs.(\ref{rep1}, \ref{rep2}, \ref{rep3}) we calculate the contribution $O(\epsilon;\mathbf{24_S})$ to the light neutrino mass matrix given in the Table \ref{tab:1}

\begin{align}\nonumber
\mathcal{L}=& (-i)\mathbf{\bar{5}}_{a}\mathbf{5}^{\phi_{u_{b}}} \mathbf{24}^{a}_{b}+(-i)\mathbf{24}^{l}_{m}\Sigma^{c}_{l}\mathbf{24}^{m}_{c}\\ \nonumber
\nonumber =& (i)^{2}\mathbf{\bar{5}}_{a}\mathbf{5}^{\phi_{u_{b}}}[\mathbf{24}^{a}_{b}\mathbf{24}^{l}_{m}]_{s}\Sigma^{c}_{l}\mathbf{24}^{m}_{c}\\ \nonumber
\nonumber =& (i)^{2}\mathbf{\bar{5}}_{5}\mathbf{5}^{\phi_{u_{5}}}[\mathbf{24}^{5}_{5}\mathbf{24}^{l}_{m}]_{s}\Sigma^{c}_{l}\mathbf{24}^{m}_{c}\\ \nonumber
\nonumber =& (i)^{2}\mathbf{\bar{5}}_{5}\mathbf{5}^{\phi u_{
5}}[\frac{5}{2}(\delta^{5}_{m}\delta^{l}_{5} + \delta^{5}_{l}\delta^{m}_{5}) - \delta^{5}_{5}\delta^{l}_{m}]\Sigma^{c}_{l}\mathbf{24}^{m}_{c}\\ \nonumber
=&(i)^{2}\mathbf{\bar{5}}_{5}\mathbf{5}^{\phi_{u_{5}}}[\frac{5}{2}\Sigma^{c}_{5}\mathbf{24}^{5}_{c} + \frac{5}{2}\Sigma^{c}_{5}\mathbf{24}^{5}_{c} - \Sigma^{c}_{m}\mathbf{24}^{m}_{c}]\\ \nonumber
 =&(i)^{2}\mathbf{\bar{5}}_{5}\mathbf{5}^{\phi_{u_{5}}}[5\Sigma^{c}_{5}\mathbf{24}^{5}_{c} - \Sigma^{c}_{m}\mathbf{24}^{m}_{c}]\\ \nonumber
 =&(i)^{2}\mathbf{\bar{5}}_{5}\mathbf{5}^{\phi_{u_{5}}}[5\Sigma^{1}_{5}\mathbf{24}^{5}_{1} - \Sigma^{1}_{1}\mathbf{24}^{1}_{1} - \Sigma^{1}_{2}\mathbf{24}^{2}_{1} - \Sigma^{1}_{3}\mathbf{24}^{3}_{1} - \Sigma^{1}_{4}\mathbf{24}^{4}_{1} - \Sigma^{1}_{5}\mathbf{24}^{5}_{1}\\ \nonumber
 +&5\Sigma^{2}_{5}\mathbf{24}^{5}_{2} - \Sigma^{2}_{1}\mathbf{24}^{1}_{2} - \Sigma^{2}_{2}\mathbf{24}^{2}_{2} - \Sigma^{2}_{3}\mathbf{24}^{3}_{2} - \Sigma^{2}_{4}\mathbf{24}^{4}_{2} - \Sigma^{2}_{5}\mathbf{24}^{5}_{2}\\ \nonumber
 +&5\Sigma^{3}_{5}\mathbf{24}^{5}_{3} - \Sigma^{3}_{1}\mathbf{24}^{1}_{3} - \Sigma^{3}_{2}\mathbf{24}^{2}_{3} - \Sigma^{3}_{3}\mathbf{24}^{3}_{3} - \Sigma^{3}_{4}\mathbf{24}^{4}_{3} - \Sigma^{3}_{5}\mathbf{24}^{5}_{3}\\ \nonumber
 +&5\Sigma^{4}_{5}\mathbf{24}^{5}_{4} - \Sigma^{4}_{1}\mathbf{24}^{1}_{4} - \Sigma^{4}_{2}\mathbf{24}^{2}_{4} - \Sigma^{4}_{3}\mathbf{24}^{3}_{4} - \Sigma^{4}_{4}\mathbf{24}^{4}_{4} - \Sigma^{4}_{5}\mathbf{24}^{5}_{4}\\ \nonumber
 +&5\Sigma^{5}_{5}\mathbf{24}^{5}_{5} - \Sigma^{5}_{1}\mathbf{24}^{1}_{5} - \Sigma^{5}_{2}\mathbf{24}^{2}_{5} - \Sigma^{5}_{3}\mathbf{24}^{3}_{5} - \Sigma^{5}_{4}\mathbf{24}^{4}_{5} - \Sigma^{5}_{5}\mathbf{24}^{5}_{5}]\\ \nonumber
 =&(i)^{2}\mathbf{\bar{5}}_{5}\mathbf{5}^{\phi_{u_{5}}}[-(2)(2)-(2)(2)-(2)(2)-(-3)(-3)+5(-3)(-3)-(-3)(-3)] \mathbf{N}_{24}\\ \nonumber
 =&(i)^{2}\mathbf{\bar{5}}_{5}\mathbf{5}^{\phi_{u_{5}}}[-4 -4 -4 -9 +45 -9]\mathbf{N}_{24}\\ \nonumber
 =&-15\alpha^{2}\mathbf{\bar{5}}_{5}\mathbf{5}^{\phi_{u_{5}}}\mathbf{N}_{24}.\\ \nonumber
\end{align} 

The contribution is $-15$ which is listed on one of the entries in the Table \ref{tab:1}. Similarly are calculated the others.

\bibliographystyle{h-physrev4}
\bibliography{tesis}

\end{document}